\documentclass[letterpaper]{aa501}

\usepackage{rotating}
\usepackage{graphicx}

\def\deg{\hbox{$^\circ$}}
\def\arcmin{\hbox{$^\prime$}}

\def\fdg{\hbox{$.\!\!^\circ$}}
\def\farcm{\hbox{$.\mkern-4mu^\prime$}}

\def\emptybox#1{\setbox1=\hbox{#1}\hbox to\wd1{}}
\def\hi{H\,{\sc i~}}

\def\NH{$N_{\rm HI}$}

\def\kms{km\,s$^{-1}$}
\def\vlsr{$V_{\rm LSR}$}

\def\vdev{$V_{\rm DEV}$}
\begin{document}
% \thesaurus{10(09.03.1; 10.11.1; 11.01.1; 11.09.4; 11.12.1)}

\title{An all--sky study of compact, isolated high--velocity clouds}

\titlerunning{An all--sky study of CHVCs}

\author{
  V. de\,Heij\inst{1},
  R. Braun\inst{2},
  and W.\,B. Burton\inst{1,3}
}

\institute{
  Sterrewacht Leiden,
    P.O. Box 9513,
    2300 RA Leiden,
    The Netherlands \and
  Netherlands Foundation for Research in Astronomy,
    P.O. Box 2,
    7990 AA Dwingeloo,
    The Netherlands \and
   National Radio Astronomy Observatory, 520 Edgemont Road,
Charlottesville, Virginia 22903, U.S.A. }

\date{Received mmddyy/ accepted mmddyy}

\offprints{R. Braun\\
\email{rbraun@astron.nl}}

\abstract {We combine the catalog of compact high--velocity \hi clouds
extracted by de\,Heij et al. (\cite{deheij02}) from the
Leiden/Dwingeloo Survey in the northern hemisphere with the catalog
extracted by Putman et al. (\cite{putman02a}) from the Parkes HIPASS
data in the southern hemisphere, and analyze the all--sky properties of
the ensemble. Compact high--velocity clouds are a subclass of the
general high--velocity cloud phenomenon which are isolated in position
and velocity from the extended high--velocity Complexes and Streams
down to column densities below 1.5$\times10^{18}$\,cm$^{-2}$. Objects
satisfying these criteria for isolation are found to have a median
angular size of less than one degree.  We discuss selection effects
relevant to the two surveys; in particular the crucial role played by
obscuration due to Galactic H\,{\sc i}.  Five principal observables are
defined for the CHVC population: \,(1) the spatial deployment of the
objects on the sky, (2) the kinematic distribution, (3) the number
distribution of observed \hi column densities, (4) the number
distribution of angular sizes, and (5) the number distribution of \hi
linewidth.  Two classes of models are considered to reproduce the
observed properties.  The agreement of models with the data is judged
by extracting these same observables from simulations, in a manner
consistent with the sensitivities of the observations and explicitly
taking account of Galactic obscuration.  We show that models in which
the CHVCs are the \hi counterparts of dark--matter halos evolving in
the Local Group potential provide a good match to the observables. The
best--fitting populations have a maximum HI mass of $10^7\rm\;M_\odot$,
a power-law slope of the HI mass distribution in the range $-1.7$
to~$-1.8$, and a Gaussian dispersion for their spatial distributions of
between 150 and 200~kpc centered on both the Milky Way and M\,31. Given
its greater mean distance, only a small fraction of the M\,31
sub--population is predicted to have been detected in present
surveys. An empirical model for an extended Galactic halo distribution
for the CHVCs is also considered. While reproducing some aspects of the
population, this class of models does not account for some key
systematic features of the population.  }

\maketitle

% ----------------------------------------------------------------------
% INTRODUCTION
% ----------------------------------------------------------------------
\section{Introduction}
Since the discovery of the \hi high--velocity clouds by Muller et
al. (\cite{muller63}), different explanations, each with its own
characteristic distance scale, have been proposed. It is likely that
not all of the anomalous--velocity \hi represents a single phenomenon,
in a single physical state.  Determining the topology of the entire
population of anomalous--velocity \hi is not a simple matter, and the
task is all the more daunting to carry out on an all--sky basis because
of disparities between the observational survey material available from
the northern and southern hemispheres.  The question of distance
remains the most important, because the principal physical parameters
depend on distance: mass varying as $d^2$, density as $d^{-1}$, and
linear size directly as $d$. Most of the \hi emission at anomalous
velocities is contributed from extended complexes containing internal
sub-structure but embedded in a common diffuse envelope, with angular
sizes up to tens of degrees.  Such structures include the Magellanic
Stream of debris from the Galaxy/LMC interaction and several HVC
complexes, most notably complexes A, C, and H.  The complexes are few
in number but dominate the \hi flux observed.

The Magellanic Stream comprises gas stripped from the Large Magellanic
Cloud, either by the Galactic tidal field or by the ram--pressure of
the motion of the LMC through the gaseous halo of the Galaxy.  It
therefore will be located at a distance comparable to that of the
Magellanic Cloud, i.e. some $50\rm\;kpc$ (see e.g. Putman \& Gibson
\cite{putm99}).  The distance to Complex A has been constrained by van
Woerden et al. (\cite{woer99}) and then more tightly by Wakker
(\cite{wakker01}) to lie within the distance range $8<d<10$ kpc.  If,
as seems plausible, the other large complexes also lie at distances
ranging from several to some 50 kpc, they will have been substantially
affected by the radiation and gravitational fields of the Milky Way.

Another category of anomalous \hi high--velocity clouds are the
compact, isolated high--velocity clouds discussed by Braun \& Burton
(\cite{braun99}).  CHVCs are distinct from the HVC complexes in that
they are sharply bounded in angular extent at very low column density
limits, i.e. below 1.5$\times$10$^{18}$\,cm$^{-2}$ (de\,Heij et
al. \cite{deheij02}). This is an order of magnitude lower than the
critical \hi column density of about 2$\times$10$^{19}$\,cm$^{-2}$,
where the ionized fraction is thought to increase dramatically due to
the extragalactic radiation field. For this reason, these objects are
likely to provide their own shielding to ionizing radiation. Although
not selected on the basis of angular size, such sharply bounded objects
are found to be rather compact, with a median angular size of less than
1~degree.

An analogy of the CHVC ensemble with that of the dwarf galaxy
population in the Local Group is suggestive, and illustrates the
hypothesis that is under discussion here.  Some few Local Group dwarf
galaxies also extend over large angles.  The Sgr Dwarf Spheroidal
discovered by Ibata et al.  (\cite{ibat94}) spans some $40\deg$;
presumably it was once a rather conventional dwarf, but its current
proximity to the Milky Way accounts for its large angular size. This
proximity has fundamentally distorted its shape, and will determine its
further evolution.  The streams of stars found in the halo of the
Galaxy by Helmi et al. (\cite{helmi99}) probably represent even more
dramatic examples of the fate which awaits dwarf galaxies which
transgress into the sphere of the Galaxy's dominance.  Analogous
stellar streams have been found in M\,31 by Ibata et al.  (\cite{ibat01})
and by Choi et al. (\cite{choi02}), as well as in association with
Local Group dwarf galaxies by Majewski et al.  (\cite{maje00}),
indicating that accretion (and subsequent stripping) of satellites is
an ongoing process.  If a selection were to be made of the dwarf galaxy
population in the Local Group on the basis of angular size, the few
large--angle systems which would be selected, namely the LMC and SMC,
and the Sgr dwarf, and --- depending on the flexibility of the selection
criteria --- perhaps the ill--fated coherent stellar streams in
the Galactic halo, would represent systems nearby, of large angular
extent, and currently undergoing substantial evolution.  Those systems
selected on the basis of being compact and isolated, on the other
hand, would represent dwarf galaxies typically at substantial
distances and typically at a more primitive stage in their evolution.
Regarding distance and evolutionary status they would differ from the
nearby, extended objects, although at some earlier stage the
distinction would not have been relevant.

The possibility that some of the high--velocity clouds might be
essentially extragalactic has been considered in various contexts by,
among others, Oort (\cite{oort66}, \cite{oort70}, \cite{oort81}),
Verschuur (\cite{vers75}), Eichler (\cite{eich76}), Einasto et al.
(\cite{eina76}), Giovanelli (\cite{giov81}), Bajaja et al.
(\cite{baja87}), Wakker \& van Woerden (\cite{wakker97}), Braun \&
Burton (\cite{braun99}), and Blitz et al. (\cite{blitz99}).  It is
interesting to note that the principal earlier arguments given against
a Local Group deployment, most effectively in the papers cited above by
Oort and Giovanelli, were based on the angular sizes of the few large
complexes and on the predominance of negative velocities in the single
hemisphere of the sky for which substantial observational data were
then available. The more complete data available now, however, show
about as many features at positive velocities as at negative ones.  It
is also interesting to note that the papers by Eichler and by Einasto
et al. cited above consider distant high--velocity clouds as possible
sources of matter, including dark matter, fueling continuing evolution
of the Galaxy.  Blitz et al. (\cite{blitz99}) revived the suggestion
that high--velocity clouds are the primordial building blocks fueling
galactic growth and evolution, and argued that the extended complexes
owe their angular extent to their proximity.

Braun \& Burton (\cite{braun99}) identified CHVCs as a subset of the
anomalous--velocity gas that might be characteristic of a single class
of HVCs, whose members plausibly originated under common circumstances
and share a common subsequent evolutionary history. They emphasized the
importance of extracting a homogenous sample of independently confirmed
objects from well--sampled, high--sensitivity \hi surveys.  The spatial
and kinematic distributions of the CHVCs were found by Braun \& Burton
to be consistent with a dynamically cold ensemble spread throughout the
Local Group, but with a net negative velocity with respect to the mean
of the Local Group galaxies.  They suggested that the CHVCs might
represent the low--circular--velocity dark matter halos predicted by
Klypin et al.  (\cite{klyp99}) and Moore et al. (\cite{moor99}) in the
context of the hierarchical structure paradigm of galactic evolution.
These halos would contain no, or only a few, stars; most of their
visible matter would be in the form of atomic hydrogen.  Although many
of the halos would already have been accreted into the Galaxy or M\,31,
some would still populate the Local Group, either located in the far
field or concentrated around the two dominate Local Group galaxies.
Those passing close to either the Milky Way or M\,31 would be
ram--pressure stripped of their gas and tidally disrupted by the
gravitational field. Near the Milky Way, the tidally distorted
features would correspond to the high--velocity--cloud complexes
observed.
 
The quality and quantity of survey material is important to
interpretation of the CHVC population, which is a global one.  The
observational data entering this analysis involved merging two catalogs
of CHVCs, one based on the material in the Leiden/Dwingeloo Survey
(LDS) of Hartmann \& Burton (\cite{hartmann97}), and the other based on
the \hi Parkes All--Sky Survey (HIPASS) described by Barnes et
al. (\cite{barnes}). Both of these surveys were searched for
anomalous--velocity features using the algorithm described by de\,Heij
et al. (\cite{deheij02}, Paper~I).  This algorithm led to the LDS
catalog of de\,Heij et al. for the CHVCs at declinations north of
$-30\deg$, and to the HIPASS catalog of Putman et
al. (\cite{putman02a}) for those at $\delta<0\deg$. The surveys overlap
in the declination range $+2\deg <\delta < -30\deg$, allowing estimates
of the relative completeness of the catalogs.  We are able to predict
how a survey with the LDS parameters would respond to the CHVCs
detected by HIPASS, and vice versa. In the subsequent simulations, we
are able to {\emph {sample the simulated material as if it were being
observed}} by one of these surveys.

This paper is organized as follows. We describe the application of the
algorithm in \S\,\ref{sec:algorithm}. In \S\,\ref{sec:selection} we
discuss various observational selection effects, and indicate how to
account for these. We address obscuration by our own Galaxy in
\S\,\ref{sec:obscur}, the consequences of the differing observational
parameters of the LDS and HIPASS data in \S\,\ref{sec:obspara}, and the
resulting differing degrees of completeness of the LDS and HIPASS
catalogs in \S\,\ref{sec:complete}.  We discuss the observable all--sky
properties of the CHVC ensemble in \S\,\ref{sec:allsky}, including the
spatial deployment (in \S\,\ref{sec:spatial}), the kinematic deployment
(in \S\,\ref{sec:velocity}), and the distributions of \hi flux and
angular size (in \S\,\ref{sec:over}).  We then attempt to reproduce
these properties by considering models, first based on Local Group
distributions as discussed in \S\,\ref{sec:model} and then on
distributions within an extended Galactic Halo as discussed in
\S\,\ref{sec:simplemodel}; these simulations are sampled {\emph {as if
being observed}} with the LDS and HIPASS programs. We discuss the
conclusions that can be drawn from this analysis in
\S\,\ref{sec:conclusion}.

\begin{figure*}
\sidecaption 
 \caption{Fraction of a homogeneously distributed sample of test clouds
that is not blended with Galactic emission.  The sample was distributed on
the sky with a Gaussian velocity distribution in the Local Standard of Rest
system (upper panel), the Galactic Standard of Rest system (middle),
and  the Local Group Standard of Rest (lower). The average velocity
of the Gaussian velocity distributions is $-50\rm\;km\;s^{-1}$ for each
of the simulations; the velocity dispersion is $240\rm\;km\;s^{-1}$ for
the LSR representation, and $110\rm\;km\;s^{-1}$ for the GSR and LGSR
ones, in rough accordance with the observed situation. The obscuration
by Galactic emission was simulated by removing that part of the sample
for which the deviation velocity (measured with respect to the LSR) is
less than $70\rm\;km\;s^{-1}$, based on a model of Galactic kinematics
which incorporates observed properties of the gaseous disk, including
its warp and flare.  The apparent structures are a consequence of the
non--uniform obscuration in position and velocity. The appearance of the
\hi Zone of Avoidance differs when material is considered in different
reference frames.}
\label{fig:calcskydistr}
\end{figure*}

\begin{figure*}
\centering
\caption{Demonstration of the influence of Galactic obscuration on
observable properties of a CHVC population. The left--hand panels show
the distributions of measured (after obscuration) versus actual mean
velocity and velocity dispersion determined from a series of 1000
simulations of a population of 200 objects.  The upper panel on the
left represents a population with a Gaussian velocity distribution in
the GSR reference frame, with $\mu = -50\rm\;km\;s^{-1}$ and $\sigma =
115\rm\;km\;s^{-1}$; the lower left--hand panel represents a population
with a Gaussian velocity distribution in the LGSR frame, with $\mu =
-55\rm\;km\;s^{-1}$ and $\sigma = 105\rm\;km\;s^{-1}$.  Obscuration
removes about 30\% of the population and leads to both an overestimate
of the dispersion and a more negative estimate of the mean velocity.
The right--hand panels illustrate the degree to which populations in the
GSR and LGSR frames could be distinguished via their statistical
parameters. Measured and actual parameter differences between the GSR
and LGSR frames are contrasted for 1000 simulated populations of 200
objects, half defined in the GSR frame and half in the LGSR frame.  All
populations have a mean velocity of~$-50\rm\;km\;s^{-1}$ and a
dispersion of~$110\rm\;km\;s^{-1}$ in their reference frame.  The
measured parameter differences for the observed CHVC sample are indicated
by the dashed lines. While the mean velocity does not provide significant
distinguishing capability between the GSR and LGSR frames, the velocity
dispersion does.
}\label{fig:obscure}
\end{figure*}

\begin{figure*}
\centering
% {\includegraphics[width=17cm]{fig3.png}} 
\caption{~Demonstration of the effects of differing resolution and
sensitivity in the LDS and HIPASS data on the extracted  CHVC
samples.~~{\it upper left:} \,Comparison of the velocity FWHM for the
CHVCs found in the HIPASS catalog (black line) with the velocity widths
of those found in the LDS catalog  (red line).  The velocity resolution
of the LDS is 1.03 \kms, but 26 \kms~for HIPASS; after degrading the LDS
data to the HIPASS velocity resolution, the dashed red line is
obtained.~~{\it upper right:} \,Comparison of the angular FWHM for the
CHVCs in the HIPASS catalog  (black line) with those in the LDS catalog
(dashed line). The angular resolution of the LDS is
$36\arcmin$, but $15\arcmin$ for HIPASS; the dashed black line shows
the HIPASS values after convolving with the LDS beam.~~{\it lower
left:} \,Comparison of the total flux of CHVCs found in the HIPASS
catalog (black line) with those in the LDS catalog (red
line).  After compensating the HIPASS detection rates for the lower LDS
sensitivity, the dashed black histogram is obtained.  The compensation
was performed using the relative detection rates in the region of
survey overlap (see Fig.~\ref{fig:completeness}).~~{\it lower
right:} \,Comparison of the total fluxes for the semi--isolated objects
(:HVCs and ?HVCs) entered in the HIPASS catalog (black line) with those
in the LDS listing (red line).  Compensation of the HIPASS detections to
the corresponding LDS sensitivity gives the dashed black line.
}\label{fig:unity}
\end{figure*}

\begin{figure*} 
\sidecaption 
\caption{Indication of the degree of completeness of the CHVC catalog
extracted from the LDS by de\,Heij et al. (\cite{deheij02}).  The solid
curve shows the fraction of external  galaxies with the indicated peak
\hi brightness temperatures that were shown by Hartmann \& Burton
(\cite{hartmann97}) to have been detected in the LDS.  Dashed lines show
the expected completeness for sensitivities that are 25\% better or worse,
respectively, than that of the LDS. The histogram indicates the fraction
of HIPASS sources from the catalog of Putman et al. (\cite{putman02a})
within each temperature range that are also found in the LDS, in the
declination zone   $-30\deg<\delta<0\deg$ where the two surveys overlap.
}\label{fig:completeness}
\end{figure*}

\begin{figure*}
\centering
% {\includegraphics[width=16cm]{fig5.png}}
\caption{~Spatial deployment of CHVCs over the sky.~~{\it upper
panel:} \,Distribution of the cataloged CHVCs, with triangles
representing the LDS sample of de\,Heij et al. (\cite{deheij02}) at
$\delta >0\deg$ and diamonds representing the HIPASS sample of Putman
et al.  (\cite{putman02a}) at southern declinations. Filled circles
correspond to the Local Group galaxies listed by Mateo (\cite{mateo}).
Red symbols indicate positive LSR velocities and black symbols negative
velocities. The background grey--scale shows \hi column depths from an
integration of observed temperatures over velocities ranging from
$V_{\rm LSR} = -450\rm\;km\;s^{-1}$ to $+400\rm\;km\;s^{-1}$, but
excluding all gas with $V_{\rm DEV} < 70\rm\;km\;s^{-1}$.~~{\it lower
panel:} \,Smoothed relative density field of the CHVCs, accounting for
the different observational parameters of the LDS and HIPASS catalogs.
The cataloged CHVCs are each represented by a Gaussian with a true--angle
dispersion of $20^\circ$; the total flux of the Gaussian is set to unity
for the LDS objects and to the likelihood of observing such an object in
an LDS--like survey for the HIPASS sources. The grey-scale is calibrated
in object number per steradian. Contours are drawn at relative densities
of $-$60\%, $-$30\%, 0\%  (in white) and 30\%, 60\%, 90\% (in black). A
significant over--density of CHVCs in the southern hemisphere remains
after accounting for the different observational parameters.
}\label{fig:skydistr}
\end{figure*}

\begin{figure*}
\sidecaption
% {\includegraphics[width=12cm]{fig6.png}}
\caption{Kinematic deployment of CHVCs identified in the LDS
(triangles) and in the HIPASS (diamonds) data, plotted against Galactic
longitude for three different kinematic reference frames, namely the
LSR (upper), the GSR (middle), and the LGSR (lower panel).  The filled
circles show the kinematic deployment with longitude of the Local Group
galaxies listed by Mateo (\cite{mateo}).  The mean velocities and the
dispersions in velocity of the CHVCs and Local Group galaxies are listed
in Table~\ref{table:velostat} for the three reference frames.
}\label{fig:glonv}
\end{figure*}

\begin{figure*}
\sidecaption
% {\includegraphics[width=12cm]{fig7.png}}
\caption{Kinematic deployment of CHVCs identified in the LDS
(triangles) and in the HIPASS (diamonds) data, plotted against Galactic
latitude in the three different kinematic reference frames, as in
Fig.~\ref{fig:glonv}.  The filled circles show the kinematic deployment
with latitude of the Local Group galaxies listed by
Mateo (\cite{mateo}). The mean velocities and the dispersions in velocity
of the CHVCs and Local Group galaxies are listed in
Table~\ref{table:velostat} for the three reference frames.
}\label{fig:glatv}
\end{figure*}

\begin{figure*}
\centering
% {\includegraphics[width=17cm]{fig8.png}}
\caption{Smoothed distributions of velocity and velocity dispersion of
the CHVC ensemble.  The panels on the left show the average velocity in
the LSR (upper), GSR (middle), and LGSR (lower) reference frames,
respectively.  The panels on the right show the velocity dispersions,
similarly arranged.  Individual CHVCs in the ensemble were convolved
with a Gaussian of true--angle dispersion of~$20^\circ$.  White
contours for the velocity and dispersion fields are at drawn at values
of $0,\;50,\;\ldots\ \rm km\;s^{-1}$; black ones are drawn at
$-50,\;-100,\;\ldots\ \rm km\;s^{-1}$.  These smoothed representations
of the observed situation can be compared with similarly sampled and
smoothed representations of simulations, as described in the text.}
\label{fig:vfield}
\end{figure*}

\begin{figure*}
\centering
% {\includegraphics[width=17cm]{fig9.png}}
\caption{Summary of the observed spatial, kinematic, angular size, and
flux properties of the CHVC ensemble.  The three panels arranged across
the top of the figure show sky projections, as follows: {\it left:}
Smoothed density field of the CHVC population.  A Gaussian with a
dispersion of~$20^\circ$ (true angle) was drawn at the location of each
CHVC; the volume of the Gaussian is unity for both LDS and HIPASS sources
-- thus in this case the observations are shown directly, i.e. HIPASS
sources are \emph {not} weighted by the likelihood with which they would
be observed in a LDS--like survey.  {\it middle:} Smoothed velocity
field of the population in the Galactic Standard of Rest frame.  {\it
right:} Smoothed velocity dispersion field.  The grey--scale bar for
the left--hand panel is labeled in units of CHVC per steradian; the
other two bars are labeled in units of \kms.  Contours are drawn at
relative densities of $-$60\%, $-$30\%, 0\% (in white) and 30\%, 60\%,
90\% (in black). White contours for the velocity and dispersion fields
are at drawn at values of $0,\;50,\;\ldots\ \rm km\;s^{-1}$; black ones
are drawn at $-50,\;-100,\;\ldots\ \rm km\;s^{-1}$.  The two panels in
the middle row of the figure show the kinematic distribution of the
observed CHVC ensemble, representing $V_{\rm GSR}$ plotted against $l$
and $b$, as indicated. Delta functions at the observed coordinates were
convolved with a Gaussian with an angular dispersion of~$20^\circ$ and
velocity dispersion of~$20\rm\;km\;s^{-1}$.  The two lower panels show,
respectively, the observed peak \hi column density distribution of the
CHVC population and the observed angular size distribution.
}\label{fig:dataoverview}
\end{figure*}

\begin{figure*}
\sidecaption
% {\includegraphics[width=12cm]{fig10.png}}
\caption{~Variation of heliocentric velocity versus the cosine of the
angular distance between the solar apex and the direction of
the object; CHVCs from the de\,Heij et al. (\cite{deheij02}) LDS
compilation at $\delta > 0\deg$ are plotted as triangles; those from
the Putman et al. (\cite{putman02a}) HIPASS compilation, as
diamonds. The CHVCs with $b~<~-65^\circ$ are plotted in red.
Local Group galaxies, from the review of Mateo~(\cite{mateo}), are
indicated by filled circles. The solid line represents the solar motion
of $V_\odot = 316\rm\;km\;s^{-1}$ towards $l=93^\circ,\;b=-4^\circ$ as
determined by Karachentsev \& Makarov (\cite{karachentsev96}).
Dashed lines give the  $1\sigma(V)$~envelope
($\pm60\rm\;km\;s^{-1}$, following Sandage \cite{sandage86}) encompassing
most galaxies firmly established as members of the Local Group. }
\label{fig:vhel}
\end{figure*}

% ----------------------------------------------------------------------
% DATA
% ----------------------------------------------------------------------
\section{Observations representing CHVCs over the entire
sky}\label{sec:data}

\subsection{Identification criteria}\label{sec:algorithm}

A full description of the cloud extraction algorithm is given in
Paper~I; the most salient aspects of the algorithm are the following:
\begin{itemize}
\item All pixels in the HIPASS and LDS material above the $1.5 \sigma$
level (as appropriate to the particular survey) were assigned to a
local intensity maximum; each pixel was assigned to the same maximum as
its brightest neighboring pixel. If the local maximum was brighter
than 3~$\sigma$, then the local maximum and the pixels
which were assigned to it were considered to constitute a cloudlet.
\item Adjacent cloudlets were merged into clouds if the brightest
enclosing contour for the two cloudlets either exceeds 40\%~of the
brighter peak or if the brightness exceeds 10$\sigma$.
\item Those merged cloudlets for which the peak temperature exceeds
5~$\sigma$ were deemed clouds, and were entered in the
catalog pending further consideration of their deviation velocity, as
described below.  The value of the noise that was used for this
selection was determined locally, whereas for the other steps a preset
noise value was used.
\end{itemize}

To minimize the influence of noise on the peak detection, the data were
first smoothed along both the spatial and spectral axes. The {\sc RMS}
fluctation level noted above refers to that within the smoothed data.
Once the relevant pixels were assigned to a cloud, unsmoothed data were
used to determine the cloud properties.

The Paper~I search algorithm led to the identification of
sub--structure within extended anomalous--velocity cloud complexes as
well as to the identification of sharply bounded, isolated sources.
Because of the importance to the present analysis of selecting only
isolated objects, we comment on the determination of the degree of
isolation. In order to determine the degree of isolation of the clouds
that were found by the algorithm, velocity--integrated images were
constructed of $10^\circ$~by~$10^\circ$ fields centered on each general
catalog entry.  The range of integration in these moment maps extended
over the velocities of all of the pixels that were assigned to a
particular cloud.  The CHVC classification then depended on the column
density distribution at the lowest significant contour level (about
3$\sigma$) of 1.5$\times$10$^{18}$\,cm$^{-2}$. We demanded that this
contour satisfy the following criteria: (1) that it be closed, with its
greatest radial extent less than the $10\deg$ by $10\deg$ image size;
and (2) that it not be elongated in the direction of any nearby
extended emission. Since some subjectivity was involved in this
assessment, two of the authors (VdH and RB) each independently carried
out a complete classification of all sources in the HIPASS and LDS
catalogs.  Identical classification was given to about 95\% of the
sample, and consensus was reached on the remaining~5\% after
re--examination.

A slightly different criterion for isolation was employed by Putman et
al. \cite{putman02a} in their analysis of the HIPASS sample of
HVCs. Rather than employing the column density contour at a fixed minimum
value to make this assessment, they employed the contour at 25\% of the
peak \NH\ for each object. Since the majority of detected objects are
relatively faint, with a peak column density near 12$\sigma$, the two
criteria are nearly identical for most objects. Only for the brightest
$\sim$10\% of sources might the resulting classifications differ. We
have reclassified the entire HIPASS HVC catalog with the absolute \NH\
criteria above, and have determined identical classifications for 1800 of
the 1997 objects listed. Given that the agreement in classification is
better than 90\%, we have chosen to simply employ the Putman et
al. classifications in the current study. In this way the analysis
presented here can be reproduced from these published sources.

High--velocity clouds are recognized as such by virtue of their
anomalous velocities. Although essentially any physical model of these
objects would predict that they also occur at the modest velocities
characteristic of the conventional gaseous disk of the Milky Way, at such
velocities the objects would not satisfy our criterion for isolation.
In our analysis, only anomalous--velocity objects with a deviation
velocity greater than~$70\rm\;km\;s^{-1}$ were considered.  As defined
by Wakker (\cite{wakker90}), the deviation velocity is the smallest
difference between the velocity of the cloud and any Galactic velocity,
measured in the Local Standard of Rest reference frame, allowed by a
conventional kinematic model in the same direction.  The kinematic and
spatial properties of the conventional Galactic~HI were described by a
thin gaseous disk whose properties of volume density, vertical
scale--height, kinetic temperature, and velocity dispersion, remain
constant within the solar radius; at larger galactocentric distances
the gaseous disk flares and warps, as described in Paper~I, following
Voskes \& Burton (\cite{voskes}).  The gas exhibits circular rotation
with a flat rotation curve constant at $220\rm\;km\;s^{-1}$. Synthetic
\hi spectra were calculated for this model Galaxy, and then deviation
velocities were measured from the extreme--velocity pixels in these
spectra for which the intensity exceeded~$0.5\rm\;K$.  Selection
against objects at $V_{\rm dev}< 70$ \kms~in the LSR frame introduces
systematic effects, as discussed in the following subsection.

Although compactness was not explicitly demanded of these isolated
objects, the 67 CHVCs found in the LDS survey and the 179 CHVCs found in
the HIPASS data have a small median angular size, amounting to   less
than $1^\circ$ FWHM.

\subsection{Selection effects and completeness}\label{sec:selection}

The CHVC samples used in this analysis were not extracted from a single,
homogeneous set of data, nor are they  free of selection effects, nor
are they complete.  We discuss below how we attempt to recognize and,
insofar as possible, to account for some inevitable limitations.
 
\subsubsection{Systematic consequences of obscuration by \hi in the Milky
Way}\label{sec:obscur}
 
The inevitable obscuration that follows from our perspective immersed
in the gaseous disk of the Milky Way, and which motivates the use of
the deviation velocity, will discriminate against some CHVC
detections. We may extend an analogy of the optical Zone of Avoidance
to the 21--cm regime.  The optical Zone of Avoidance refers to
extinction of light by dust in the Milky Way, and thus traces a band
with irregular borders but roughly defined by $|b|<5^\circ$; kinematics
are irrelevant in the optical case, since the absorption is
broad--band. The HI searches for galaxies in the optical Zone of
Avoidance carried out by, among others, Henning et al. (\cite{henn98})
in the north, and by Henning et al. (\cite{henn00}) and Juraszek
(\cite{jura00}) in the south, were confined to $|b|<5\deg$.

In the 21--cm regime, extinction due to high--optical--depth foreground
\hi is largely negligible, but confusion due to line blending occurs at
all latitudes. The analogous \hi zone refers to a certain range in
velocity, of varying width depending on $l$ and on $b$, but present to
some extent everywhere: near zero LSR velocity, the ``\hi Zone of
Avoidance" covers the entire sky. The nearby LSB galaxy
Cep~I was discovered during CHVC work (Burton et al. \cite{burt99});
although it is at a relatively substantial latitude, $b=8\fdg0$, its
velocity of $V_{\rm hel}= 58$ \kms~locates it within the
\hi~obscuration zone.  Because of the strong dependence on velocities
measured with respect to the Local Standard of Rest, the zone of
obscuration is distorted upon transformation to a different kinematic
reference frame.  (We note that the Magellanic Stream and the HVC
complexes plausibly also discriminate against CHVC detections, but
because these extended features are smaller in scale and more confined
in velocity than the Galaxy, we do not consider them further here.)

The relationship between the different velocity reference systems used
to characterize the CHVC kinematics is given by the equations below.
\begin{equation}\label{eqn:vlsr}
v_{\rm LSR}=v_{\rm HEL}+9\cos(l)\cos(b)+12\sin(l)\cos(b)+7\sin(b)
\end{equation}
\begin{equation}
v_{\rm GSR}=v_{\rm LSR}+0\cos(l)\cos(b)+220\sin(l)\cos(b)+0\sin(b)
\end{equation}
\begin{equation}
v_{\rm LGSR}=v_{\rm GSR}-62\cos(l)\cos(b)+40\sin(l)\cos(b)-35\sin(b)
\end{equation}
Note that a typographical error is present (the sign of the coefficient
of $sin(b)$) in the version of Eqn.\ref{eqn:vlsr} that is published in
Braun \& Burton (\cite{braun99}).

The influence of the obscuration by the modeled Galaxy is illustrated
by Fig.~\ref{fig:calcskydistr}, where the integral $\int\exp(-(V -
\mu)^2 / 2\sigma^2) dV$ is plotted. The range of integration extends
over all velocities which deviate more than~$70\rm\;km\;s^{-1}$ from
any Local Standard of Rest (LSR) velocity allowed by the Galactic model
described above; $\mu$ is the average velocity and $\sigma$ is the
standard deviation of the test clouds.  The panels in
Fig.~\ref{fig:calcskydistr} show the fraction of a population of
clouds, homogeneously distributed on the sky and with a Gaussian
velocity distribution relative to a particular reference frame, that
are not obscured by virtue of being coincident with \hi emission from
the Milky Way.  The upper panel of the figure represents a model in
which the Gaussian velocity distribution is with respect to the Local
Standard of Rest frame, with an average velocity of
$-50\rm\;km\;s^{-1}$ and dispersion of $240\rm\;km\;s^{-1}$, in rough
agreement with the measured CHVC values in this frame. In this case the
obscuration is simply proportional to the velocity width of the
obscuring emission. The obscuration at high latitudes is quite uniform
since the infalling population is always displaced from \vlsr=0 \kms,
where the obscuring gas resides.  The middle panel presents a model
wherein the Gaussian velocity distribution is with respect to the
Galactic Standard of Rest (GSR) frame with an average velocity of
$-50\rm\;km\;s^{-1}$ and dispersion of $110\rm\;km\;s^{-1}$, in rough
agreement with the CHVC values in this frame. The low--latitude
obscuration is similar to that in the LSR model, although more strongly
modulated since the velocity dispersion is smaller. The high--latitude
obscuration is quite strongly modulated since the infall velocity in
the GSR frame overlaps with \vlsr=0 \kms~in the plane approximately
perpendicular to the direction of rotation,
$(l,b)~=~(90^\circ,0^\circ)$.  Broad apparent maxima in unobscured
object density are centered near $l=90\deg$ and $l=270\deg$.  The lower
panel presents a model in which the Gaussian velocity distribution is
with respect to the Local Group Standard of Rest (LGSR), again using an
average velocity of $-50\rm\;km\;s^{-1}$ and dispersion of
$110\rm\;km\;s^{-1}$. The pattern of obscuration is very similar to
that of the GSR case, although the maxima in unobscured object density
are slightly shifted with respect to $b=0\deg$. These results indicate
that caution must be exercised in interpreting apparent spatial
concentrations of detected objects without properly accounting for the
distortions introduced by the \hi Zone of Avoidance.

We have also considered how the measured statistics of a distribution,
namely  the mean velocity and dispersion, are influenced by the
non--completeness caused by obscuration.  Figure~\ref{fig:obscure} shows
the distribution of the errors in the average velocity and dispersions
for 1000~simulations, each involving 200~test clouds; one set of
simulations was run with the GSR as the natural reference frame, and a
second set was run with the LGSR as the natural frame.  After removing
the test clouds that have an LSR deviation velocity less
than~$70\rm\;km\;s^{-1}$, the velocity dispersion of the simulated
ensemble was measured for both the GSR and the LGSR velocity systems,
and compared with what would have been determined if there had been no
obscuration by the Galaxy.

The upper left--hand panel in Fig.~\ref{fig:obscure} refers to test
objects with a Gaussian distribution in $V_{\rm GSR}$ with a dispersion
of~$115\rm\;km\;s^{-1}$ and average of~$-50\rm\;km\;s^{-1}$.  The
measured dispersion exceeds the true one by~$9\rm\;km\;s^{-1}$, whereas
a more negative average velocity is inferred by~$12\rm\;km\;s^{-1}$.
The lower left--hand panel is based upon test samples with a Gaussian
distribution in $V_{\rm LGSR}$ with a dispersion
of~$105\rm\;km\;s^{-1}$ and an average of~$-55\rm\;km\;s^{-1}$.  The
differences between the measured and true dispersion and average
velocity, of $6\rm\;km\;s^{-1}$ and $-5\rm\;km\;s^{-1}$, respectively,
are smaller than for the GSR system.  From the 200~clouds which were in
the input ensemble, an average of 80 were removed because of
obscuration in the GSR model and only~60 in the LGSR model, indicating
that the statistical properties of the LGSR model are somewhat better
preserved in this case. The particular population attributes chosen
above for the GSR and LGSR systems were chosen to match the observed
parameters in these systems, as shown below in \S\,\ref{sec:allsky}.

Another question that can be addressed with these simulations is
whether it might be possible to distinguish between a GSR and an LGSR
CHVC population based on a significant difference in the
statistical properties. We assessed this by taking 500 populations of
200 objects in both the GSR and LGSR frames, each with a dispersion
of 110~\kms~and an average velocity of~$-$50~\kms. Each of these 1000
populations was analyzed in both the GSR and LGSR frames, both before
and after decimation by obscuration. The results are shown in the
right--hand panels of Fig.~\ref{fig:obscure} for differences in velocity
dispersion (relative to the GSR versus LGSR frames) and mean velocity,
respectively. The measured differences in velocity dispersion and mean
velocity of our CHVC sample (from \S\,\ref{sec:allsky}) are plotted in
these panels as  dashed lines.  The model results for mean velocity
differences form a continuous cloud, for which it is impossible to
distinguish between the actual reference frame of the model
population. The model results for velocity
dispersion differences, on the other hand, are separated into two
distinct clouds. The velocity dispersion of each model population is
minimized in its own reference frame with a variance of only a
few~\kms, while the dispersions within the GSR and LGSR frames are
separated by about 20~\kms, both before and after obscuration. The
measured difference in velocity dispersion of the CHVC sample relative
to the GSR and LGSR frames, of 16~\kms, is more consistent with an LGSR
reference frame. 

\subsubsection{Consequences of the differing observational parameters of
the LDS and HIPASS}\label{sec:obspara}

Because the LDS and the HIPASS data do not measure the sky with the
same limiting sensitivities, angular resolutions, velocity resolutions,
or velocity coverages, the northern population of CHVCs will be
differently sampled than the southern one.  In particular, the maximum
depth of the two samples will be different since the surveys have
different limiting fluxes. We describe below how we identify, and
compensate for, the differing properties of the two catalogs; we also
describe how we sample the simulations using the selection criteria
corresponding to the observations. A detailed comparison of objects
detected in the two surveys is made in Paper~I.

The LDS covered the sky north of declination $-30^\circ$ (the actual
declination cut-off varied between $-32$ and $-28^\circ$); the
angular resolution of the 25--m Dwingeloo telescope was
$36^\prime$. The effective velocity coverage of the LDS extends over
LSR velocities from $-450\rm\;km\;s^{-1}$ to~$+400\rm\;km\;s^{-1}$,
resolved into channels $1.03\rm\;km\;s^{-1}$~wide.  The formal rms
sensitivity is~0.07~K per 1.0~\kms~channel. Stray radiation has been
removed as described by Hartmann et al. (\cite{hartmann96}). Due to the
presence of radio frequency interference, it was important that the
reality of all CHVC candidates that were identified in the LDS be
independently confirmed.  Although interference in the LDS often had
the shape of extremely narrow--band signals that are easily recognized
as artificial, some types of interference were indistinguishable from
naturally occurring features. The reality of the CHVC candidates was
either confirmed by the identification of the candidates with objects
in independent published material, or by new observations made with the
Westerbork Synthesis Radio Telescope, operating as a collection of 14
single dishes.

The HIPASS program covered the sky south of declination~$+2^\circ$.
The  survey has been reduced in such a way that emission which
extends over more than~$2^\circ$ was filtered out.  To recover a larger
fraction of the extended emission, the part of the survey which covers
LSR velocities ranging from~$-700\rm\;km\;s^{-1}$
to~$+500\rm\;km\;s^{-1}$ was re--reduced using the {\sc minmed5} method
described by Putman (\cite{putman}), before production of the Putman et
al. (\cite{putman02a}) catalog.  The HIPASS data were gridded with
lattice points separated by $4^\prime$ with an angular resolution of
$15\farcm5$. The HIPASS velocity resolution after Hanning smoothing
is~$26.4\rm\;km\;s^{-1}$, thus substantially coarser than the 1.03
\kms~of the LDS.  The HIPASS sensitivity for such a velocity resolution
is 10~mK for unresolved sources.  Because the observing procedure
involved measuring each line of sight five times in order to reach the
full sensitivity, all HIPASS sources have effectively been confirmed
after median gridding.

Figure~\ref{fig:unity} shows that the LDS and the HIPASS reflect
differing measures of the CHVC properties, because of their differing
observational properties.  The panel in the upper left of this figure
contrasts the observed velocity widths of the LDS and HIPASS
samples. The velocity FWHM  measured in the LDS  ranges from about
20~\kms~to some 40~\kms, with a median of about 25~\kms.  Only for a
few sources were values as low as 5~\kms~measured.  The relatively
high median FWHM likely indicates that most of the observed \hi in the
CHVCs is in the form of warm neutral medium.  High--resolution
observations of a sample of ten CHVCs made with the $3\farcm5$
resolution afforded by the Arecibo telescope (Burton et al.
\cite{burt01}) showed warm halos to be a common property of these
objects. On the other hand, the median HIPASS velocity width is about
35~\kms~FWHM. We can demonstrate that the two observed FWHM
distributions are consistent with the same object population by
convolving the LDS distribution with the HIPASS velocity resolution.
The resulting distribution agrees well with that measured in the
HIPASS. 

The panel in the upper right of Fig.~\ref{fig:unity} shows histograms
of the angular sizes of the cataloged CHVCs, determined from velocity
integrated images of each cloud. A contour was drawn at the intensity
of half the peak column density of the cloud.  After fitting an ellipse
to this contour, the size of the cloud was measured as the average of the
minor and major axes.  It is clear from these distributions that many
of the CHVCs are resolved by HIPASS, but that this is rarely the case
for the LDS.  Some CHVCs in the LDS catalog were only detected in a
single spectrum --- giving the peak in the histogram at~$0\fdg4$,
which is an artifact of the sub--Nyquist LDS sky sampling.
After convolving the HIPASS distribution with the LDS beam a
more similar distribution of sizes is found, although there remains
a small excess of relatively large objects in the north.

The panel on the lower left of Figure~\ref{fig:unity} shows the
flux distribution for the CHVCs detected by HIPASS and the LDS,
respectively.  An excess of faint sources is present in the HIPASS
sample, even after compensation for the lower LDS sensitivity (as outlined
below). Conversely, the LDS may have a small excess of bright objects.
If semi-isolated objects are considered (i.e. the :HVC and ?HVC categories
discussed by Putman et al.  \cite{putman02a}  and de\,Heij et
al.  \cite{deheij02})  as in the   panel on the lower right of
Fig.~\ref{fig:unity}, these differences remain, with the adjusted
HIPASS sample showing an excess of faint sources in the south and the LDS
sample showing a small excess of brighter sources in the north.

 \subsubsection{Completeness and uniformity of the
CHVC samples}\label{sec:complete}

The finite sensitivity of the LDS and HIPASS observations results in
sample incompleteness at low flux levels in both surveys. The different
sensitivities of the two surveys will bias the derived
sky--distribution, average velocity, and velocity dispersion towards
the more sensitively observed hemisphere, namely the southern one.  To
compensate for this bias, the objects found in the southern hemisphere
were weighted with the likelihood that they would be detected by a
survey with the LDS properties.  For this likelihood we use the
relation plotted in Fig.~\ref{fig:completeness}, following de\,Heij et
al. (\cite{deheij02}), who assess the degree of completeness of the LDS
catalog as a function of limiting peak brightness from a comparison of
the detection rates of cataloged external galaxies over the range
$-30^\circ<\delta<90^\circ$, and from a comparison with the HIPASS
catalog of Putman et al. (\cite{putman02a}) for the range
$-30^\circ<\delta<0^\circ$.  To incorporate plausible uncertainties in
this relation, the calculations have also been done for a fictional
survey~25\% more sensitive, and for one 25\% less sensitive, than the
LDS, as indicated by the dashed lines in the figure.

Table~\ref{table:nsource} lists the number of sources with a minimum
peak brightness temperature for the northern hemisphere, as observed by
the LDS, and for the southern one, as observed by HIPASS.  Due to the
differences in spectral and spatial resolution, the LDS and HIPASS
measure different peak temperatures for the same cloud.  For all clouds
that are observed in both surveys, the median of the temperature ratio
as measured in HIPASS and LDS is~1.5 (de\,Heij et al. \cite{deheij02}).
Applying this temperature scaling to the HIPASS data provides very good
agreement with the external galaxy completeness curve of
Fig.~\ref{fig:completeness} for declinations $-30\deg$ to $0\deg$.
However, over the entire HIPASS declination range, the compensated
HIPASS data show a strong excess in the source detection rate for
sources with an LDS peak temperature in the range 0.2 to 0.4~K.
According to Fig.~\ref{fig:completeness}, the LDS completeness for
these sources should exceed~80\%. Therefore the difference in the
numbers of relatively faint CHVCs detected by HIPASS and LDS indicates
an asymmetry in the distribution upon the sky, with about a factor of
two more occurring in the southern hemisphere than in the
north. Reducing the sensitivity of the LDS survey by~25\% does not
change this conclusion.

The CHVC tabulation is probably not incomplete as a consequence of the
velocity--range limits of the observational material.  Although the
part of the LDS that was searched only extended over the range $-450 <
V_{\rm LSR} < +350$ \kms, de\,Heij et al.  (\cite{deheij02}) plausibly
did not miss many (if any) clouds because of this limited interval.
The high--velocity feature with the most extreme negative velocity yet
found is that discovered by Hulsbosch (\cite{hulsbosch78}) at
\vlsr$=-466$\,\kms.  (This object is listed in Paper~I as
?HVC\,$110.6\!-\!07.0\!-\!466$: being incompletely sampled in velocity,
it does not meet the stringent isolation criteria for the CHVC
category, and so does not enter this analysis further.) The Wakker \&
van Woerden (\cite{wakk91c}) tabulation, which relied on survey data
covering the range $-900 < V_{\rm LSR} < +750$ \kms, found no
high--velocity cloud at a more negative velocity.  The HIPASS search by
Putman et al. (\cite{putman02a}) sought anomalous--velocity emission
over the range $-700 < V_{\rm LSR} < +1000$ \kms.  Of the HIPASS CHVCs
cataloged by Putman et al., only 10 have $V_{\rm LSR}<-300$, but the
most extreme negative velocity is $-353$ \kms, for
CHVC\,$125.1\!-\!66.4\!-\!353$.  Regarding the positive--velocity
extent of the ensemble, we note that only 7 objects in the HIPASS
catalog have $V_{\rm LSR}$ greater than $+300$ \kms, and only one has a
velocity greater than $350$ \kms, namely
CHVC\,$258.2\!-\!23.9\!+\!359$.  All of the 7 CHVCs with substantial
positive velocities are near $(l,b)~=~(270^\circ,0^\circ)$, where
Galactic rotation contributes to a high positive LSR velocity. Since
this extended region has a negative declination, it is sampled with the
wider velocity coverage of HIPASS, rather than that of the LDS.  In
view of these detection statistics, we consider it unlikely that the
velocity--range limits of either the LDS or of the HIPASS have caused a
significant number of CHVCs to be missed.  In other words, the true
velocity extent, as well as the non--zero mean in the LSR frame, of the
anomalous--velocity ensemble are well represented by the observed
extrema of $-466$ \kms~and $+359$ \kms.

The strong concentration of faint CHVCs with an extreme variation in
their radial velocity in the direction of the south Galactic pole was
already noted by Putman et al. (\cite{putman02a}). A complete model for
the all-sky distribution of objects will need to reproduce the
enhancement in numbers as well as local velocity dispersion in this
direction. Much of the north--south detection asymmetry for faint CHVCs
remains even after excluding all objects with a Galactic latitude less
than~$-65^\circ$, as we discuss in detail below.

\begin{table*}
\caption{Number of sources with a minimum peak temperature detected
in the northern hemisphere and listed in the LDS catalog of de\,Heij et
al. (\cite{deheij02}), and in the southern hemisphere and listed in the
HIPASS catalog of Putman et al. (\cite{putman02a}).  Because of the
differing angular and velocity resolutions, the two surveys measure
different peak temperatures for the same source.  The median ratio of
HIPASS to LDS peak temperature of 1.5 determined for the sources in common
 has been used to resample the HIPASS data in the last column.
}\label{table:nsource}

\begin{tabular}{c|ccc}
  \hline
\vspace{-.05cm} minimum & $ N_ {\rm LDS}$ & $  N_{\rm HIPASS}$ & $
N_{\rm HIPASS}$
\\
$T_{\rm peak}\rm\;[K]$ & $>T_{\rm peak}$ & $>T_{\rm peak}$ &
$>1.5\,T_{\rm peak}$ \cr
  \hline        
      1.0 &         3 &         5 &          3 \cr
          0.5 &         9 &        24 &          9 \cr
          0.4 &        12 &        37 &         16 \cr
          0.3 &        20 &        56 &         29 \cr
          0.2 &        30 &        85 &         56 \cr
          0.1 &        38 &       160 &        115 \cr
  \hline
\end{tabular}

\end{table*}

\begin{figure*}
\sidecaption
% {\includegraphics[width=12cm]{fig11.png}}
\caption{~Average velocity field in the Local Group entering the
simulations described in \S\,\ref{sec:model}.  The velocity at each
grid point is given by the average velocity of all the test particles
located in a box centered on the grid point and with a width of 10~kpc.
Squares are drawn if the velocity dispersion of the ensemble of
particles exceeds $100\rm\;km\;s^{-1}$.  The length of the thick line
in the upper left corresponds to a velocity of~$200\rm\;km\;s^{-1}$.
The image corresponds to a simulation with a Local Group mass of
$4.3\times10^{12}\rm\;M_\odot$.  The Milky Way and M\,31 are located at
$(x = -0.47, y = 0.0 {\rm\;Mpc})$ and $(x = 0.23,
y = 0.0 {\rm\;Mpc})$, respectively.  The contours show the relative
density levels of a combination of two Gaussian distributions with
200~kpc dispersion, centered on the Milky Way and M\,31 at 1, 5, 10,
20, 40, and 80\% of the peak.  }
\label{fig:modelfield}
\end{figure*}

\begin{figure*}
\sidecaption
% {\includegraphics[width=12cm]{fig12.png}}
\caption{~Properties of the CHVCs entering the simulations described in
\S\,\ref{sec:model}.  Plotted as a function of \hi mass, the images
show the FWHM of the \hi distribution, the central \hi volume density,
the velocity dispersion of the gas, and the peak column
density. Details of the relation between the \hi masses and cloud
properties depend on the dark-matter fraction via the power-law slope
of the \hi mass distribution of the CHVC population being modeled: the
dashed lines in the images correspond to a slope $\beta=-1.2$; the
solid lines, to a slope of~$-1.6$; and the dotted lines, to a slope
of~$-2.0$.  }
\label{fig:chvcmodel}
\end{figure*}

\begin{figure*} 
\sidecaption 
\caption{ Distances out to which simulated CHVCs would be detected in
the HIPASS survey.  The relation between the \hi masses and maximum
observable distance depends on the dark--matter fraction via the slope
of the \hi mass distribution of the CHVC population.  The three curves
refer to clouds with a mass--distribution slope of $-2.0, -1.6$, and
$-1.2$, as dotted, solid, and dashed lines, respectively. The
$\beta = -1.2$ clouds are so diffuse that they fall below the HIPASS
detection threshold for  log($M_{\rm HI}) < \sim6.4$.  The horizontal
lines bracket distances to individual Sculptor Group galaxies.  }
\label{fig:obsdistance}
\end{figure*}

\begin{figure*} 
\sidecaption 
\caption{ Distances up to which simulated CHVCs of the indicated total
\hi masses would be destroyed by ram--pressure stripping in the
Galactic halo. As an illustration, the limiting distance is calculated
assuming a relative velocity of~$200\rm\;km\;s^{-1}$.  The relation
between the \hi masses and the stripping distance depends on their
dark--matter content via the slope of the \hi mass distribution of the
CHVC population.  The dotted curve corresponds to
a slope of $\beta = -2.0$, the solid curve to $\beta = -1.6$, and the
dashed one to a slope
of~$-1.2$. } \label{fig:ramdistance}
\end{figure*}

\begin{figure*} 
\sidecaption 
\caption{Tidal force for different potentials as a function of
Galactocentric distance. The dashed line corresponds to a point source
with the mass of the Milky Way, the dotted line corresponds to a
potential consistent with a rotation curve flat at the level
of~$220\rm\;km\;s^{-1}$, while the solid line corresponds to the
isochrone potential which is used in the simulations to describe the
Milky Way.  The dashed horizontal lines correspond to the indicated \hi
mass slopes for a cloud mass of $M_{\rm HI}=10^5\rm\;M_\odot$. Only
for this low cloud mass and the low dark--matter fraction implied by
$\beta=-2$ are clouds unstable to tidal disruption. }
\label{fig:tidalfield}
\end{figure*}

\begin{figure*}
\centering
% {\includegraphics[width=16cm]{fig16.png}}
\caption{ Demonstration of the effects of shot--noise on fit quality,
showing the best-- and worst--fitting instances from a sequence of 35
simulations with one of the lowest average $\chi^2$ values. The
parameter values are $M_1~=~10^7$~M$_\odot$, $\beta~=~-1.7$, and
$\sigma_d~=~200$~kpc, corresponding to  model \#9 from
Table~\ref{table:bestfit}. The best--fitting instance is plotted on the
right and the worst--fitting on the left. Black lines and symbols are
used for the simulations and red for the observations. Values of $\chi^2$
from top to bottom for the best-fitting case are 1.5, 2.2, 3.2, 0.19, and
0.21, respectively; while for the worst-fitting case these are 3.5, 3.3,
5.5, 0.29, and 0.36.  }\label{fig:qualmodel}
\end{figure*}

\begin{figure*}
\centering
% {\includegraphics[width=17cm]{fig17.png}}
\caption{Overview of the spatial and kinematic properties of one of the
best fitting Local Group models from the simulations of
\S\,\ref{sec:model}.  The simulation, model~\#9 in
Table~\ref{table:bestfit}, has the following parameters: $M_1 =
10^7\rm\;M_\odot$, $\beta = -1.7$, and $\sigma_d = 200\rm\;kpc$.  The quality
of the fit to the various observables is given by $\chi^2 ({\rm size})
= 2.3$, $\chi^2 ({\rm N_{HI}}) = 2.9$, $\chi^2 ({\rm FWHM}) = 4.1$,
$\chi^2 (l,b) = 0.28$, $\chi^2 (l,V_{\rm GSR}) = 0.24$, and $\chi^2
(V_{\rm GSR},b) = 0.29$. The panels provide the same information as
the panels in Fig.~\ref{fig:dataoverview} for the observed data.  The
simulation was sampled with the observational parameters of the LDS and
HIPASS surveys, depending on the declination of the test cloud, as
discussed in the text. The thick--line histograms indicate the LDS
(northern hemisphere) contributions to the total
detections. }\label{fig:model01}
\end{figure*}

\begin{figure*}
\centering
% {\includegraphics[width=17cm]{fig18.png}}
\caption{Overview of the spatial and kinematic properties of one of the
best fitting Local Group models from the simulations of
\S\,\ref{sec:model}.  The simulation, model~\#3 in
Table~\ref{table:bestfit}, has the following parameters: $M_1 =
10^{7.5}\rm\;M_\odot$, $\beta = -1.7$, and $\sigma_d = 150\rm\;kpc$.
The quality of the fit is characterized by $\chi^2 ({\rm size}) = 2.6$,
$\chi^2 ({\rm N_{HI}}) = 2.7$, $\chi^2 ({\rm FWHM}) = 3.9$, $\chi^2
(l,b) = 0.25$, $\chi^2 (l,V_{\rm GSR}) = 0.21$, and $\chi^2 (V_{\rm
GSR},b) = 0.25$.  The panels provide the same information as the panels
in Fig.~\ref{fig:dataoverview} for the observed data. The thick--line
histograms indicate the LDS (northern hemisphere) contributions to the
total detections.  }\label{fig:model02}
\end{figure*}

\begin{figure*}
\centering
% {\includegraphics[width=17cm]{fig19.png}}
\caption{Overview of the spatial and kinematic properties of one of the
best fitting Local Group models before including the effects of
foreground obscuration and SGP exclusion. The simulation, model~\#\,9
in Table~\ref{table:bestfit}, is shown in Fig.~\ref{fig:model01} after
applying these effects.  The panels provide the same information as the
panels in Fig.~\ref{fig:dataoverview} for the observed
data. }\label{fig:model01no}
\end{figure*}

\begin{figure*}
\centering
% {\includegraphics[width=17cm]{fig20.png}}
\caption{Three--dimensional distribution of synthetic clouds in the
model \#9 simulation of a CHVC population in the Local Group.  The
Galaxy and M\,31 are indicated with the large black dots, with the Galaxy
at $(x,y,z)=(0,0,0)$.  The axes are labeled in units of Mpc.  The smaller
circles indicate all of the objects in the model, whose parameters are
given in Table~\ref{table:bestfit}.  Not all of the clouds survive the
simulated environment, and not all of those that do survive would be
detected in the LDS and HIPASS observations.  The filled black circles
indicated those input clouds that are destroyed by tidal and
ram--pressure stripping influences of M\,31 and the Galaxy.  The filled
grey circles indicate clouds that are too faint to be detected by the LDS
or by HIPASS, respectively, depending on their declination as viewed from
the origin.  The open red circles are the objects that are obscured by the
foreground Galactic \hi.  Only the filled red circles would be detected in
the combined LDS and HIPASS CHVC sample.  }\label{fig:mod3d}
\end{figure*}

% ----------------------------------------------------------------------
% CLOUD PROPERTIES
% ----------------------------------------------------------------------
\section{All--sky spatial, kinematic, and column density properties of
the CHVC ensemble}\label{sec:allsky}

We show in this section the basic observational data for the all--sky
properties of the CHVCs; specifically, the deployment in position and
velocity as well as the perceived size and \hi column density
distributions.  These basic properties constitute the observables
against which the simulations described in the following sections are
tested.

\subsection{Distribution of CHVCs on the sky }\label{sec:spatial}

Figure~\ref{fig:skydistr} shows the all--sky distribution of the
cataloged CHVCs superimposed on the integrated \hi emission observed in
the range $-450<$\vlsr$<+400$ \kms, but with \vdev$>70$ \kms.  The LDS
catalog and data are used in the north and the HIPASS catalog and data
in the south, with a solid line marking the demarcation at
$\delta=0^\circ$ separating the LDS from the HIPASS material.  Red
symbols indicate positive LSR velocities and black symbols negative
velocities\footnote{Several of the figures in this paper make
color-coded distinctions; although the printed journal and electronic
versions will display the color coding, black-and-white printouts will
not, of course.}  The much higher object density observed in the
southern hemisphere is quite striking, as is the absence of diffuse
emission in the HIPASS {\sc minmed5} data. We comment further below on
the extent to which the CHVC density is a consequence of the differing
observational parameters, especially that of sensitivity.

To get a better impression of the CHVC clustering and distribution on
the sky, an average density field is constructed; this smoothed field
is more appropriate for comparison with simulated fields, which, as
indicated below, are similarly smoothed.  A field of delta functions at
the CHVC locations was convolved with a Gaussian with a dispersion
of~$20^\circ$.  The total flux of each delta function is set to unity
for the LDS sources and to the value of the likelihood that such a
particular CHVC would be observed in an LDS--like survey for the HIPASS
sources.  Changes in the likelihood relation do not change the overall
picture of the CHVC concentrations; only the contrasts of the
overdensity regions with respect to the average changes.

Figure~\ref{fig:skydistr} shows that the projected density of CHVCs
displays a number of local enhancements. The three most prominent of
these occur in the southern hemisphere, and were previously noted by
Putman et al. (\cite{putman02a}) as Groups~1 through 3. Group~1 is
concentrated at the south Galactic pole and extends from about
$b=-60\deg$ to $-90^\circ$. It is remarkable for possessing a local
velocity dispersion in excess of 150~\kms, about twice that seen in any
other part of the sky. This region is bisected by a portion of the
Magellanic Stream and is also spatially coextensive with the nearest
members of the Sculptor group of galaxies (with D$\sim$1.5~Mpc).
Group~2 is located near $(l,b) \sim (280\deg,-15\deg)$, with an
extent of about $30\deg$. This concentration is approximately in the
direction of the leading arm of the Magellanic Clouds but is also near
the Local Group anti-barycenter direction, where the Blitz
et~al. (\cite{blitz99}) model predicts an enhancement of high--velocity
clouds. Group~3 is centered near $(l,b) \sim (30\deg,-15\deg)$, a
region that Wakker \& van Woerden (\cite{wakk91c}) have identified with
the GCN (Galactic Center Negative velocity) population. The most
diffuse concentration, which we label Group~4, is in the northern sky
near $(l,b) \sim (115\deg,-30\deg)$, approximately coinciding with
the Local Group barycenter.  The Blitz et~al.  (\cite{blitz99}) model
also predicts an enhancement of high--velocity clouds here, albeit a
stronger one than observed. Likewise the mini--halo simulations of
Klypin et al. (\cite{klyp99}), Moore et al. (\cite{moor99},
\cite{moor01}), and Putman \& Moore (\cite{putman02c}) predict a
strongly enhanced density of low mass objects around the major galaxies
of the Local Group, in particular toward M\,31, which lies close to the
barycenter direction.  We comment further on the expected strength of
such an enhancement in the observed distribution below.

\subsection{Distribution of CHVCs in velocity }\label{sec:velocity}

The kinematic properties of the CHVC population provide an important
constraint that must be reproduced by a successful model of the
phenomenon.  The kinematic distribution is plotted against Galactic
longitude and latitude, for the Local, Galactic, and Local Group
kinematic reference frames in Figs.~\ref{fig:glonv} and~\ref{fig:glatv},
respectively.  The CHVCs are confined within a kinematic envelope
narrower in extent than the \vlsr~spectral coverage of the surveys; we
stressed above in \S\,\ref{sec:complete} that this confinement is not a
selection effect; it is one of the global kinematic properties of the
ensemble which must be accounted for.

Table~\ref{table:velostat} shows that the ensemble of clouds has a lower
velocity dispersion in both the GSR and LGSR systems, compared to that
measured in the LSR frame, suggesting that either the Galaxy or the Local
Group might be the natural reference system of the CHVCs.  By measuring
the dispersion in the LSR frame, one introduces the solar motion around
the Galactic center into the velocities, which results in a higher
dispersion.

The CHVC groups noted in the previous subsection can also be identified
in the $(l,V)$ and $(b,V)$ distributions. Group~1 is best seen in
Fig.~\ref{fig:glatv} where it gives rise to the very broad velocity
extent in both the GSR and LGSR frames for $b<-60^\circ$. Group~2, on
the other hand, is best seen in Fig.~\ref{fig:glonv}, centered near
$l=280^\circ$. This group has a positive mean velocity in the GSR
frame. Only by going to the LGSR frame does the mean group velocity
approach zero. Group~3 is evident in both Figs.~\ref{fig:glonv}
and~\ref{fig:glatv}. This concentration is  seen near
$l = 30^\circ$ and has a remarkably high negative velocity of about
$-200$~\kms~in both the GSR and LGSR frames. Group~4 can also be
distinguished near $l=115^\circ$ in Fig.~\ref{fig:glonv}. This group
also retains a large negative velocity in both the GSR and LGSR frames.

Table~\ref{table:velostat} gives the all--sky statistical parameters of
the CHVC ensemble, calculated by weighting the HIPASS objects with
the likelihood that they would be observed in an LDS--like survey. The
variation of these parameters with the (flux--dependent) relative
weighting of the HIPASS sub--sample is explored by considering both 25\%
higher and lower relative sensitivity. Although the dispersion is not
affected strongly by the weighting given to the HIPASS sub--sample, the
mean velocity becomes increasingly negative as the fainter HIPASS
sub--sample receives a higher relative weight.

\begin{table*}
\caption{Average velocity and velocity dispersion for the CHVC
ensemble and for the dwarf galaxies in the Local Group, expressed for
three different kinematic reference frames.  To correct for the
difference in sensitivity between the LDS and the HIPASS compilations,
the HIPASS CHVCs were weighted by the likelihood that they would be
observed in an LDS--like survey. The three values given for the average
velocity and for the dispersion for the CHVC ensemble for each
reference frame, pertain to an LDS sensitivity 25\%~lower than the one
shown in Fig.~11 of Paper~I, to the same sensitivity, and to a
sensitivity that is 25\% higher, respectively.  The Local Group data
refer to 27 dwarf galaxies with known radial velocities, from the
tabulation of Mateo (\cite{mateo}). }
\label{table:velostat}

\begin{tabular}{c|cccc}
  \hline
\vspace{-.05cm} & CHVCs & CHVCs & L.G. galaxies & L.G. galaxies \\
\vspace{-.05cm} reference frame & $<$velocity$>$ & dispersion &
$<$velocity$>$
& dispersion
\\
 & (\kms) & (\kms) & (\kms) & (\kms) \cr
\hline
                  &   $-33$ &    $253$ & & \cr
  LSR             &   $-45$ &    $238$ & $-57$ & $196$\cr
                  &   $-59$ &    $240$ & & \cr
  \hline           
                  &   $-58$ &    $128$ & & \cr
  GSR             &   $-63$ &    $128$ & $-22$ & $104$\cr
                  &   $-69$ &    $126$ & & \cr
  \hline           
                  &   $-57$ &    $114$ & & \cr
  LGSR            &   $-60$ &    $112$ & $+4$ & $79$ \cr
                  &   $-65$ &    $110$ & & \cr
  \hline
\end{tabular}
\end{table*}

CHVCs near the galactic equator display the horizontal component of their
space motion.  Figure~\ref{fig:glatv} shows that the radial motions
at low $|b|$ are at least as large as those at high latitudes, and
furthermore that the CHVC distribution does not avoid the Galactic
equator, and that substantial positive--velocity amplitudes, as well as
negative--velocity ones, are observed. Large horizontal motions as well
as high positive velocities are difficult to account for in terms of a
galactic fountain model (e.g. Shapiro \& Field \cite{shap76}, Bregman
\cite{breg80}).  Similarly, CHVCs  located near the galactic poles offer
unambiguous information on the vertical, $z$, component of their space
motion. The vertical motions are substantial, with positive velocities
approximately equal in number and amplitude to negative velocities; the
vertical motions are of approximately the same amplitude as the
horizontal ones.  This situation also is incompatible with the
precepts of the fountain model, which predicts negative $V_z$ velocities
for material returning in a fountain flow. Furthermore, the values of
$V_z$ are predicted to not exceed the velocity of free fall, of some 200
\kms. In fact, \vlsr~amplitudes substantially larger than the
free--fall value are observed.

Several aspects of the spatial and kinematic topology of the class are
difficult to account for if the CHVCs are viewed as a Milky Way
population, in particular if they are viewed as consequences of a
galactic fountain; these same aspects would seem to discourage a
revival of several of the mechanisms suggested earlier for a Milky Way
population of high--velocity clouds (reviewed, for example, by Oort
\cite{oort66}), including ejection from the Galactic nucleus,
association with a Galactic spiral arm at high latitude, and ejection
following a nearby supernova explosion.  We note that the spatial
deployment plotted Fig.~\ref{fig:skydistr} shows no preference for the
Galactic equator, nor for the longitudes of the inner Galaxy expected
to harbor most of the disruptive energetic events.  CHVCs do not
contaminate the \hi terminal--velocity locus in ways which would be
expected if they pervaded the Galactic disk; this observation
constrains the clouds either to be an uncommon component of the Milky
Way disk, confined to the immediate vicinity of the Sun, or else to be
typically at large distances beyond the Milky Way disk.  We note also
that the lines of sight in the directions of each of the low $|b|$
CHVCs traverse some tens of kpc of the disk before exiting the Milky
Way: unless one is prepared to accept these CHVCs as boring through the
conventional disk at hypersonic speeds (for which there is no
evidence), and atypical in view of the cleanliness of the
terminal--velocity locus, then their distance is constrained to be
large.  We note further that some of the CHVC objects are moving with
velocities in excess of a plausible value of the Milky Way escape
velocity (cf. Oort \cite{oort26}).

Figure~\ref{fig:vfield} shows the average velocity field and velocity
dispersion field, which is constructed in the same way as the average
density field.  A field of delta functions was convolved with a
Gaussian with a dispersion of~$20^\circ$.  The flux of each delta
function was set equal to the measured CHVC velocity and multiplied by
the likelihood that the CHVC would be observed in an LDS--like
survey. The convolved image was then normalized by the density
field. For the velocity dispersion field, a gridded distribution of
squared velocity was similarly generated and the velocity dispersion
was calculated from the square root of the mean squared velocity less
the mean velocity squared, $\sigma=\sqrt(<V^2>-<V>^2)$. The velocity
dispersion field was blanked where the normalized density was below the
mean, since insufficient objects otherwise contribute to the measurement
of local dispersion.

Kinematic patterns in the LSR velocity field are dominated by the
contribution of Galactic rotation. After removing the contribution of
Galactic rotation by changing to the GSR reference frame, the following
characteristics of the kinematics of the groups are evident.  Relative
minima of $V_{\rm GSR} =-100$ to $-175$~\kms~are seen in the directions
of Groups~3 and 4, and a relative maximum of $V_{\rm GSR} =
+45\rm\;km\;s^{-1}$ is seen in the vicinity of Group~2. Transforming to
the LGSR frame generally lowers the magnitude of these kinematic
properties (except in the case of Group~3 which becomes more negative
in velocity) although they are all still present. The relative
velocities of Groups 2, 3, and 4 fit into a coherent global pattern
shared by much of the CHVC population, consisting of a strong gradient
in the GSR and LGSR velocity that varies from strongly negative below
the Galactic plane in the first and second quadrants to near zero in
the third and fourth quadrants near the plane.

The distribution of velocity dispersion is not as strongly effected by
the choice of reference frame since it is a locally defined quantity.
The exception to this rule is near $l=0\deg$, where there are large
gradients in the velocity field, leading to larger apparent dispersions
when sampled with our smoothing kernel.  Group~1 is remarkable for its
extremely high velocity dispersion, exceeding that of Groups~2--4 by a
factor of two or more.  It is plausible that the Group~1
concentration represents a somewhat different phenomenon than the
remainder of the CHVC sample, as we discuss further below.

\subsection{Summary of the basic observables of the CHVC
ensemble}\label{sec:over}

In the preceding sub--sections we have attempted to correct for the
differing detection levels in the northern and southern hemisphere data
to produce a spatially unbiased estimate of the CHVC distributions in
position and velocity. However, when making comparisons with model
calculations it is possible to explicitly take account of the differing
resolutions and sensitivity of the data in the north and south,
obviating the need to re--weight portions of our sample in advance. The
basic observables from our all--sky sample of CHVCs, without
re--weighting, are shown in Fig.~\ref{fig:dataoverview} relative to the
GSR frame. The top row of panels represents the density, velocity, and
velocity-dispersion fields, just as in Figs.~\ref{fig:skydistr} and
\ref{fig:vfield}, except that the HIPASS sub--sample has not been
re-weighted relative to the LDS.  Smoothed versions of the $(l,V)$ and
$(V,b)$ plots shown in Figs.~\ref{fig:glonv} and~\ref{fig:glatv} are
shown in the middle panels of Fig.~\ref{fig:dataoverview} to facilitate
comparison with the model distributions discussed below. The
distribution of delta functions was convolved with a Gaussian with a
dispersion of~$20^\circ$ in angle and 20~\kms\ in velocity.  Composite
histograms of the peak column density and angular size distributions
for the whole sky are shown in the lower panels of
Fig.~\ref{fig:dataoverview}. Since the LDS and HIPASS survey
resolutions are different (as discussed above in \S\,\ref{sec:obspara})
these observables have a different physical implication in the two
hemispheres, but again, these differences can be accounted for
explicitly in the comparison with model distributions.

\begin{figure*}
\sidecaption
% {\includegraphics[width=12cm]{fig21.png}}
\caption{~Velocities plotted against galactic longitude for the
ensemble of synthetic clouds corresponding to model \#\,9 of a CHVC
population in the Local Group, whose sky deployment is plotted in
Fig.~\ref{fig:model01}. The red symbols indicate unobscured clouds
which are sufficiently bright to be detected by the Leiden/Dwingeloo or
Parkes surveys, depending on the object declination. Black symbols
refer to simulated clouds which are either too faint to be detected or
obscured.  As in the observed velocity, longitude plots of
Fig.~\ref{fig:glonv}, the kinematic distributions for the simulated
situation are indicated for three different kinematic reference frames,
namely the LSR (upper), the GSR (middle), and the LGSR (lower
panel). }\label{fig:mod01glonv}
\end{figure*}

\begin{figure*}
\sidecaption
% {\includegraphics[width=12cm]{fig22.png}}
\caption{~Velocities plotted against galactic latitude for the ensemble
of synthetic clouds corresponding to model \#\,9 of a CHVC population
in the Local Group. The kinematic distributions are indicated for three
different kinematic reference frames, namely the LSR (upper), the GSR
(middle), and the LGSR (lower panel).  As in the simulated ($l,V$)
distribution of Fig.~\ref{fig:mod01glonv}, red symbols indicate
unobscured clouds which are sufficiently bright to be detected by the
Leiden/Dwingeloo or Parkes surveys, depending on the object
declination, while black symbols refer to simulated clouds which are
either too faint to be detected or obscured.  The simulated ($b,V$)
plot may be compared with the observed situation plotted in
Fig.~\ref{fig:glatv}. }\label{fig:mod01glatv}
\end{figure*}

\begin{figure*}
\centering
% {\includegraphics[width=17cm]{fig23.png}}
\caption{Predicted distribution on the sky of detected synthetic clouds
corresponding to the model \#9 simulation of a CHVC population in the
Local Group, contrasting LDS and HIPASS sensitivities; the parameters
of the simulation are given in Table~\ref{table:bestfit}.  The black
dots correspond to objects that exceed the LDS and HIPASS detection
threshold and are not obscured by Galactic H\,{\sc i}. The red dots are
predicted additional detections if the HIPASS sensitivity were extended to
the northern hemisphere.  The superposed grey--scale image shows, as in
Fig.~\ref{fig:skydistr}, the observed \hi column depths, following from
an integration of observed temperatures over velocities ranging from
$V_{\rm LSR} = -450\rm\;km\;s^{-1}$ to
$+400\rm\;km\;s^{-1}$, but excluding all gas with $V_{\rm DEV} <
70\rm\;km\;s^{-1}$. The boundary between the LDS regime at $\delta >
0\deg$ and the HIPASS regime at lower declinations is evident in the
grey--scale image. A much higher CHVC concentration, relative to that
currently observed in Fig.~\ref{fig:skydistr}, is predicted in the
direction of the Local Group barycenter at this increased sensitivity.
}\label{fig:mod01hom}
\end{figure*}

\begin{figure*}
\sidecaption
% {\includegraphics[width=12cm]{fig24.png}}
\caption{~Demonstration of the quality of the fits of the
Galactic Halo models discussed in \S\,\ref{sec:simplemodel}. The
upper panels show the column--density fits; the lower panels show the
size fits.  The panels on the left represent one of the
better fits, characterized by $\chi^2 ({\rm size}) = 1.7$ and $\chi^2
({\rm N_{HI}}) = 0.7$; the panels on the right show one of the poorer
fits in the acceptable category, characterized by $\chi^2 ({\rm size})
= 4.9$ and $\chi^2 ({\rm N_{HI}}) = 4.9$.  The observed distributions are
indicated by the red histograms, against which the simulations are judged.
}\label{fig:qualsimplemodel}
\end{figure*}

\begin{figure*}
\centering
% {\includegraphics[width=17cm]{fig25.png}}
\caption{Summary of the simulated spatial and kinematic properties of
the CHVC ensemble, for one of the better fits of the empirical Galactic
Halo model described in \S\,\ref{sec:simplemodel}. The panels give the
properties of the simulation in the same way as the panels in
Fig.~\ref{fig:dataoverview} describe the observed data. This Galactic
Halo simulation is determined by the following parameters: $M_0 =
10^4\rm\;M_\odot$, $n_0 = 3\times10^2\rm\;cm^{-3}$, $\beta = -2.0$, and
$\sigma_d = 30\rm\;kpc$.  The quality of the fit is described by
$\chi^2 ({\rm size}) = 2.3$ and $\chi^2 ({\rm N_{HI}}) = 2.6$. The
thick--line histograms indicate the LDS (northern hemisphere)
contributions to the total detections.  }\label{fig:abestsimplemodel}
\end{figure*}

\begin{figure*}
\centering
% {\includegraphics[width=17cm]{fig26.png}}
\caption{Summary of the simulated spatial and kinematic properties of
the CHVC ensemble, for one of the better fits of the empirical Galactic
Halo model described in \S\,\ref{sec:simplemodel}. The panels give the
properties of the simulation in the same way as the panels in
Fig.~\ref{fig:dataoverview} describe the observed data.  This Halo
simulation is determined by the following parameters: $M_0 =
10^5\rm\;M_\odot$, $n_0 = 3\times10^{-3}\rm\;cm^{-3}$, $\beta = -1.4$,
and $\sigma_d = 200\rm\;kpc$.  The quality of the fit is described by
$\chi^2 ({\rm size}) = 2.9$ and $\chi^2 ({\rm N_{HI}}) = 2.5$.  The
thick--line histograms indicate the LDS (northern hemisphere)
contributions to the total detections. Although this Halo model is
fundamentally different from the Local Group models, the characteristic
distance of the simulated CHVCs is similar.
}\label{fig:bbestsimplemodel}
\end{figure*}

\begin{figure*} 
\sidecaption 
\caption{ Histogram of object distances for one of the best-fitting
Local Group models. The detected objects in model\#\,9 of
Table~\ref{table:bestfit} are plotted in the histogram. A broad peak in
the distribution extends from about 200 to 450~kpc, with outliers as
far as 1~Mpc. The filled circles along the top of the figure are the
distance estimates for individual CHVCs of Braun \& Burton
(\cite{braun00}) and Burton et al. (\cite{burt01}), based on several
different considerations. While few in number, they appear consistent
with this distribution. }
\label{fig:mod01dist}
\end{figure*}

\begin{figure*} 
\sidecaption 
\caption{ Comparison of a model Local Group CHVC population with the
population of Local Group galaxies. The distribution of \hi masses of
one of our best--fitting Local Group models, model \#9 of
Table~\ref{table:bestfit}, is plotted as a thin--line histogram after
accounting for ram--pressure and tidal stripping and as a thick line
after also accounting for Galactic obscuration and the finite
sensitivities of the LDS and HIPASS observations. The Local Group
galaxies (excluding M31 and the Galaxy) tabulated by Mateo
(\cite{mateo}) are plotted as the hatched histogram after summing the
\hi mass with the stellar mass assuming $M/L_B
=3$\,M$_\odot$/L$_\odot$. The diagonal line has the slope of the \hi
power--law mass function, $\beta=-1.7$.  }
\label{fig:lgall}
\end{figure*}

\begin{figure*} 
\sidecaption 
\caption{ External appearance of a model Local Group CHVC population.
The spatial distribution of one of our best--fitting Local Group models,
namely model \#9 of Table~\ref{table:bestfit}, is shown projected onto a
plane as in Fig~\ref{fig:mod3d}. The grey
dots correspond to objects which have been disrupted by ram--pressure or
tidal stripping during a close passage. The red and black dots
correspond to are objects which should survive these processes, with the
color of the dot distinquishing objects by \hi mass.  The red dots
indicate clouds more massive than
$M_{\rm HI}$=3$\times10^6$\,M$_\odot$; the black dots indicate CHVCs
falling below this mass limit.  }
\label{fig:mod01mass}
\end{figure*}

% ----------------------------------------------------------------------
% A LOCAL-GROUP CHVC MODEL
% ----------------------------------------------------------------------
\section{A Local Group population model for the CHVC ensemble}
\label{sec:model}

Determining Local Group membership for nearby galaxies is not a trivial
undertaking. The well--established members of the Local Group have been
used to define a mass--weighted Local Group Standard of Rest,
corresponding to a solar motion of $V_\odot = 316\rm\;km\;s^{-1}$
towards $l=93^\circ,\;b=-4^\circ$ (Karachentsev \& Makarov
(\cite{karachentsev96}). The $1\sigma$ velocity dispersion of Local
Group galaxies with respect to this reference frame is about
60~\kms~(Sandage \cite{sandage86}). A plot of heliocentric velocity versus
the cosine of the angular distance between the solar apex and the galaxy
in question, as shown in Fig.~\ref{fig:vhel}, has often been used to
assess the likelihood of Local Group membership (e.g. van den Bergh
\cite{berg94}) when direct distance estimates have not been available.
Local Group galaxies tend to lie within about one sigma of the line
defined by the solar motion in the LGSR reference frame. Braun
\& Burton (\cite{braun99}) pointed out how the original LDS CHVC sample
followed this relationship, although offset with a significant infall
velocity. The all--sky CHVC sample has been plotted in this way in
Fig.~\ref{fig:vhel}. Both the sample size and sky coverage are
significantly enhanced relative to what was available to Braun \& Burton. A
systematic trend now becomes apparent in the CHVC kinematics. While the
CHVCs at negative $cos(\theta)$ (predominantly in the southern hemisphere)
tend to lie within the envelope defined by the LGSR solar apex and the
Local Group velocity dispersion, the CHVCs at positive $cos(\theta)$ have a
large negative offset from this envelope. Obscuration by Galactic \hi
may well be important in shaping this trend. Only by analyzing realistic
model populations and subjecting them to all of the selection and
sampling effects of the existing surveys can meaningful conclusions be
drawn. 

Recently several simulations have been performed to test the hypothesis
that the CHVCs are the remaining building blocks of the Local Group.
Putman \& Moore (\cite{putman02c}) compared the results of the full
N--body simulation described by Moore et al. (\cite{moor01}) with
various spatial and kinematic properties of the CHVC distribution, as
well as with properties of the more general HVC phenomenon, without
regard to object size and degree of isolation.  Putman \& Moore were
led to reject the Local Group deployment of CHVCs, for reasons which we
debate below. Blitz et al. (\cite{blitz99}) performed a restricted
three body analysis of the motion of clouds in the Local Group.  In
their attempt to model the HVC distributions, Blitz et al.  modeled the
dynamics of dark matter halos in the Local Group and found support for
the Local Group hypothesis when compared qualitatively with the
deployment of a sample of anomalous--velocity \hi containing most HVCs,
but excluding the large complexes (including the Magellanic Stream) for
which plausible or measured distance constraints are available.
Assuming that 98\%~of the Local Group mass is confined to the Milky Way
and M\,31 and their satellites, Blitz et al.  described the dynamics of
the Local Group in a straightforward manner.  Driven by their mutual
gravity and the tidal field of the neighboring galaxies, the Milky Way
and M\,31 approach each other on a nearly radial orbit.  The motion of
the dark matter halos which fill the Local Group was determined by the
gravitational attraction of the Milky Way and M\,31, and the tidal
field of the neighboring galaxies.  All halos which ever got closer
than a comoving distance of 100~kpc from the Milky Way or M\,31 center
were removed from the sample.  Blitz et al. describe how their
simulations account for several aspects of the HVC observations.  We
follow here the modeling approach of Blitz et al., but judge the
results of our simulations against the properties of the CHVC sample,
viewing the simulated data {\it as if it were observed} with the LDS
and HIPASS surveys.

\subsection{Model description}

We use a test particle approach similar to that used by Blitz et
al. (\cite{blitz99}) to derive the kinematic history of particles as a
function of their current position within the Local Group, but combine
this with an assumed functional form (rather than simply a uniform
initial space density) to describe the number density distribution of
the test particles.  The density function contains a free parameter
which sets the degree of concentration of the clouds around the Milky
Way and M\,31.  By determining a best fit of the models we are able to
constrain the values of this concentration parameter, and thereby
constrain the distance to the CHVCs.  The fits depend upon the derived
velocity field and the \hi properties of the clouds.

The density fields which we use are a sum of two Gaussian distributions,
centered on the Galaxy and M\,31, respectively.  As a free parameter we
use the radial dispersion of these distributions. This free parameter
has the same value for both the concentration around the Galaxy and that
around M\,31.  The ratio of the central densities of the distributions at
M\,31 and the Galaxy must also be specified. We set this ratio
equal to the mass ratio of the two galaxies.  The Gaussian dispersions
which are used in the models range from 100~kpc to 2~Mpc;
i.e.  the distributions range from very tightly concentrated around the
galaxies to an almost homogeneous filling of the Local Group.

The CHVC kinematics are simulated by tracking the motions of small test
particles within the gravitational field of the Milky Way, the M\,31
system, and the nearby galaxies.  Both the description of the tidal
field that is produced by the nearby galaxies, and the properties of
the Galaxy--M\,31 orbit, are taken from the analysis of Raychaudhury
\& Lynden--Bell (\cite{raychaudhury}), who  studied the influence of
the tidal field  on the motion of the Galaxy and M\,31, deriving
the dipole tidal field from a catalog of galaxies compiled by
Kraan--Korteweg (\cite{kraan}).  The motions of the Galaxy and
M\,31 are determined in a simulation. In this simulation M\,31
and the Galaxy are released a short time after the Big Bang.  The
initial conditions are tuned in such a way that the relative radial
velocity and position correspond with the values currently measured.

We track not only the motions of M\,31 and the Galaxy, but also the
motions of a million test particles.  The test particles are, together
with M\,31 and the Milky Way, released a short time after the Big Bang
with a velocity equal to that of the Hubble flow.  Initially, the test
particles homogenously fill a sphere with a comoving radius of~2.5~Mpc.
Their motions are completely governed by the gravitational field of
M\,31 and the Milky Way and by the tidal field of the nearby galaxies.
The result of the simulation at the present age of the Universe is
used to define the kinematic history as a function of current 3--D
position within the Local Group for our simulated CHVC populations.
For every 3--D position where an object is to be placed by our assumed
density distribution, we simply assign the kinematic history from the
test particle in the kinematic simulation which ended nearest to
that 3--D position.  The most important aspects of the kinematic
history are merely the final space velocity vector as well as the
parameters of the closest approach of the test particle to M\,31 and
the Milky Way, where the effects ram pressure and tidal stripping will
be assessed.

The parameters which determine the outcome of the simulation are the
Hubble constant, $H$, the average density of the Universe, $\Omega_0$,
the total mass of the Local Group, $M_{\rm LG}$, (=$M_{\rm M31}$+$M_{\rm
MW}$), and the mass--to--light ratio of the nearby galaxies, which make
up the tidal field.  The Hubble constant and the average density of the
Universe not only set the age of the Universe, but also the initial
velocities of the test particles.  Further evolution is independent of
these parameters.  The evolution is set by the values of the tidal
field and the masses of M\,31 and the Galaxy.  Values for all these
parameters were taken from Raychaudhury and Lynden--Bell
(\cite{raychaudhury}), namely: $H = 50 \rm\;km\;s^{-1}$, $M_{\rm LG} =
4.3\times10^{12}\rm\;M_\odot$, a mass--to--light ratio of~60, $\Omega_0
= 1$, and $M_{\rm M\,31} / M_{ \rm MW} = 2$.  Whereas the mutual
gravitational attraction between the Milky Way and M\,31 is described
by a {\rm point--mass potential}, the gravitational attraction of these
galaxies on the test particles is described by an isochrone potential
of the form
\begin{equation}
    \Phi_{\rm iso} (r) = - \frac{{\rm G} \; M}{r_0 + \sqrt{{r_0}^2 + r^2}},
\end{equation}
where $M$ is the total mass of the galaxy and $r_0$ is a characteristic
radius; $r_0$~is set such that the rotation velocity as derived from
the potential equals the measured one at the edge of the unwarped \hi
disk, i.e.  $V_{\rm circ}^{\rm MW} (12 \rm\;kpc) = 220\;km\;s^{-1}$ and
$V_{\rm circ}^{\rm M\,31} (16 \rm\;kpc) = 250\;km\;s^{-1}$.
Figure~\ref{fig:modelfield} shows the average velocity field superposed
on density contours for a Gaussian distribution of the test particles,
characterized by a dispersion of $200\rm\,kpc$. The ellipsoidal
turn--around surface of the Local Group can be seen where the velocity
vectors approach zero length at radii near 1.2 Mpc. The velocity field
is approximately radial at large distances from both M31 and the
Galaxy, but becomes more complex at smaller radii.

Before we can simulate the way in which a Local Group population of
clouds would be observed by a HIPASS-- or LDS--like survey, we have to
set the \hi properties of the test clouds.  To do so, we assume that
the \hi clouds are isothermal gas spheres, with each such cloud located
inside a dark--matter halo.  Given the temperature of the gas and the
potential in which it resides, the density profile follows from the
relation
\begin{equation} \label{eq:density}
    n(r) = n_0 \cdot \exp
        \left(
        - \frac{\rm {m}_{\rm HI} } { {\rm k} \, T_{\rm eff} }
        \left[
            \Phi(r) - \Phi(0)
        \right]
    \right),
\end{equation}
where $\rm{m}_{\rm HI}$ is the mass of the hydrogen atom, $\Phi(r)$ is
the potential at a distance~$r$ from the cloud center, and $T_{\rm
eff}$ is an effective gas temperature. Since in addition to the thermal
pressure there is also rotational support of a gas cloud against the
gravitational attraction of the dark--matter halo, an effective
temperature is used which incorporates both processes.  In general, the
average energy of an atom equals ~$\frac{1}{2} {\rm k}T$ per motional
degree of freedom, so we have defined the effective temperature such that
\begin{equation} \label{eq:temperature}
    \frac{3}{2} {\rm k} \, T_{\rm eff} =
    \frac{3}{2} {\rm k} \, T_{\rm kin} + \frac{1}{2} {\rm m}_{\rm HI}
V^2_{\rm circ},
\end{equation}
where $T_{\rm kin}$ is the gas kinetic temperature, taken to be 8000~K,
and $V_{\rm circ}$ is a characteristic rotation velocity.

The description of the gravitational potential of the dark matter halo
follows that of Burkert (\cite{burkert}), who  was able to fit a
{\emph {universal}} function to the rotation curves of four different
dwarf galaxies. The shape of the function is completely set by the
amount of dark matter in the core,~$M_0$.  The potential, which is
derived from the rotation curve, has the form
\begin{eqnarray*}
    \Phi(r) -\Phi(0) & = & -\pi \; {\rm G} \; \rho_0 \; r_0^2 \; \{ \\
                     &   & 2 \left( 1 + \frac{r_0}{r} \right) \cdot
                           \ln\left( 1 + r / r_0 \right) - \\
                     &   & 2 \left( 1 + \frac{r_0}{r} \right) \cdot
                           \arctan \left( r / r_0 \right) - \\
                     &   & \emptybox{2} \left( 1 - \frac{r_0}{r} \right)
                           \cdot \ln \left( 1 + (r / r_0)^2 \right) \; \},
\end{eqnarray*}
where the core radius, $r_0$, and the central density, $\rho_0$, are set by
the   relations
\begin{eqnarray*}
       r_0 & = & 3.07 \; \left( \frac{M_0}{10^9\rm\;M_\odot}
                 \right)^{3/7} \rm\;kpc \\
{\rm and} \\
    \rho_0 & = & 1.46 \cdot 10^{-24} \; \left( \frac{M_0}{10^9\rm\;M_\odot}
                 \right)^{-2/7} \rm\;g\;cm^{-3}. \\
\end{eqnarray*}
The circular--velocity rotation curve has a maximum value
\begin{equation}
   V_{\rm circ}^{\rm max} = 48.7 \; \left( \frac{M_0}{10^9 \rm M_\odot}
                   \right)^{(2/7)}\rm\;km\;s^{-1}.
\end{equation}
We use $V_{\rm circ}^{\rm max}$ as a parameter for $V_{\rm circ}$ in
Eq.~\ref{eq:temperature}.  Because the total mass corresponding to the
given potential is infinite, the dark matter mass is characterised
either by the core mass, $M_0$, or by the virialized mass of the halo,
$M_{\rm vir}$. We adopt a total dark--matter mass of $M_{\rm vir}$ for
each object. According to Burkert (\cite{burkert}), these are related
by $M_{\rm vir} = 5.8\;M_0$.

Although we could use Eq.~\ref{eq:density} directly to determine the
predicted 21--cm images of a CHVC given an \hi mass, we instead chose
to approximate the corresponding column--density distribution by a
Gaussian, in order to enable faster evaluation of the simulation.  Two
parameters specify the Gaussian, namely the central density, $n_0$, and
the FWHM, derived as follows.  The volume--density distribution given
by Eq.~\ref{eq:density}, can be closely approximated with an
exponential form with a matched scale--length, $h_e$, defined by
$n(h_e)~=~n(0)/e$. The column density distribution can then be
expressed in an analytic form, containing a modified Bessel function of
order 1 (e.g. Burton et al. \cite{burt01}).  This analytic
representation of the column--density distribution is then approximated
by a Gaussian of the same halfwidth from, FWHM$=2.543 h_e$.  Knowing
the mass of the \hi gas cloud and its FWHM, the central density of the
Gaussian can be determined.

To get the amount of \hi mass in the cloud, we adopt the relations
between the dark--matter mass and baryonic mass for galaxies. For
normal, massive galaxies there is approximately ten times as much dark
matter as there is baryonic matter.  Chiu et al. (\cite{chiu}) show
that this ratio depends on the total mass.  Whereas the mass spectrum
of the dark--matter halos in their simulation has the form $n(M_{\rm
dark}) \propto M_{\rm dark}^{-2}$, the baryonic mass spectrum has the
form $n(M_{\rm HI}) \propto M_{\rm HI}^{-1.6}$.  The difference in
slope is due to the ionizing extragalactic radiation field.  The lowest
mass halos are simply unable to retain their ionized baryonic
envelopes, which have a kinetic temperature of $10^4$K.  The slopes of
both the baryonic and the dark--matter distributions completely determine
the mass dependency of the ratio between dark matter and baryonic
matter.  Furthermore, given the fact that the ratio equals~10 for
objects with a baryonic mass of~$10^9\rm\;M_\odot$, the ratio is set
for all masses.  Although the simulation of Chiu et al.  gives a value
of~$-1.6$ for the baryonic mass spectrum slope, we explore a range of
values for the \hi mass spectrum.  The most appropriate value is then
obtained by fitting the models to the observations.

For the sake of simplicity, we assume that the baryonic mass of each
simulated cloud is entirely in the form of \hi. In fact, a significant
mass fraction will be in the form of ionized gas. It is likely that
the mass fraction of ionized gas will increase toward lower masses such
that below some limiting mass the objects would be fully ionized. A
realistic treatment of the ionized mass fraction was deemed beyond the
scope of this study. However, we do comment further on the implications
of this simplifying assumption where appropriate.

The definitions of the velocity and density fields of the test objects
in the Local Group, together with their \hi properties, resulted in
simulated CHVC populations which could be sampled with the
observational parameters of an LDS-- or HIPASS--like survey.  The free
parameters, defined above, were allowed to take the following values:
\begin{itemize}
\item The Gaussian dispersion of the density distributions
centered on the Galaxy and M\,31 can range from 100~kpc to 2~Mpc.
\item The slope of the \hi mass spectrum, $\beta$, not only sets the
number ratio of the less--massive with respect to the more--massive
objects, it also determines the dark--to--baryonic mass ratios.
The slope is allowed to range over the values $-2.0, -1.9, -1.8, -1.7,
-1.6, -1.4, -1.2, -1.0$.
\item $M_1$~is the highest \hi mass which is allowed for
clouds in the simulations.  The logarithm of this mass is allowed the
values $6.0, \;6.5,\;\ldots\;, \;9.0$. It was found necessary to
introduce this  upper mass cut--off  since otherwise high column
densities \NH~$\sim~10^{21}$\,cm$^{-2}$ were predicted, such as observed
in actively star-forming galaxies, but unlike what is found in the CHVC
population.
\end{itemize}

Figure~\ref{fig:chvcmodel} shows the basic cloud properties as a
functions of \hi mass. As indicated above, the dark--matter fraction as
function of mass is determined by $\beta$, the slope of the \hi mass
spectrum, so three curves are shown in each panel, corresponding to
$\beta = -1.2$, $-1.6$, and $-2.0$, respectively. The typical object
size and internal velocity dispersion increases only slowly with \hi
mass, from about 1.5 to 4~kpc, and 10 to 20 \kms, respectively, between
$M_{\rm HI}=10^5$ and $10^8$ M$_{\odot}$. The central \hi volume
density varies much more dramatically with \hi mass, as does the peak
column density. Note that the peak column densities of simulated clouds
only exceed N$_{\rm HI}~>~10^{19}$cm$^{-2}$ for M$_{\rm
HI}~>~10^{5.5}$~M$_\odot$ and $\beta$ in the range $-1.6$--$-2.0$. It
is critical that peak column densities of this order are achieved in
long-lived objects, since this is required for self-shielding from the
extragalactic ionizing radiation field (eg. Maloney \cite{maloney93},
Corbelli and Salpeter \cite{corbelli93}). 

In order to compare the simulation results with the observational data
in the most effective way, and thereby constrain the model parameters,
we created a single CHVC catalog from the HIPASS and LDS ones.
Thirty-eight CHVCs at $\delta \ge 0^\circ$ were extracted from the de\,Heij
et al. (\cite{deheij02}) LDS catalog, and 179 at $\delta <
0^\circ$ from the Putman et al. (\cite{putman02a}) HIPASS one. A large
concentration of faint sources (Group~1 noted above) with an
exceptionally high velocity dispersion is detected toward the Galactic
south pole. Because this overdensity may well be related to the nearby
Sculptor group of galaxies (see Putman et al. \cite{putman02a}), the 53
CHVCs at $b < -65^\circ$ were excluded from comparison with the
simulations. 

A simulation was run for each set of model parameters. We
chose a position for each object, in agreement with the spatial density
distribution of the ensemble, but otherwise randomly. The velocity is
given by the velocity field described above.  The \hi mass of the test
cloud is randomly set in agreement with the given power--law mass
distribution between the specified upper mass limit and a lower mass
limit described below. The physical size and linewidth of each object
follow from the choice of
$\beta$. Once all these parameters were set, we determined the observed
peak column density and angular size. Objects in the northern
hemisphere were convolved with a beam appropriate to the LDS, while
those at $\delta < 0\deg$ were convolved with the HIPASS beam.
Simulated objects were considered detected if the peak brightness
temperature exceeded the detection threshold of the relevant survey,
i.e. the LDS for objects at $\delta >0\deg$, and HIPASS otherwise.
Furthermore, in order for a test object to be retained as detected its
deviation velocity was required to exceed $70\rm\;km\;s^{-1}$ in the LSR
frame, and its Galactic latitude to be above $b = -65^\circ$ (for
consistency with the exclusion of Group~1 from our CHVC sample noted
above). In addition, as we describe below, each simulated cloud should
be stable against both tidal disruption and ram--pressure stripping by
the Milky Way and M\,31. We continued simulating additional objects
following this prescription until the number of detected model clouds
was equal to the number of CHVCs in our observed all--sky sample.

Before carrying out each simulation with a power--law distribution of
\hi masses we began by determining an effective lower \hi mass limit,
$M_0$, to the objects that should be considered. This was necessary to
avoid devoting most of the calculation effort to objects too faint to be
detected in any case. As a first guess we took $M_0=0.5 M_1$. A
sub-sample of twenty objects in the mass range $M_0$--$M_1$ was
simulated which were deemed stable to both disruption and stripping and
were detectable with the relevant survey parameters. Generating twenty
detectable objects typically required evaluating of order 500 test
objects. Given the total number of test objects needed to generate this
observable sub--sample and the slope of the \hi mass distribution
function, it is possible to extrapolate the number of required test
objects in other intervals of \hi mass belonging to this same
distribution. The predicted number of test objects in the interval
$0.67 M_0$ to $M_0$ was simulated. If at least one of these was
deemed detectable, then the lower \hi mass limit was replaced with
$0.67 M_0$ and the procedure outlined above was repeated. This
process continued until no detectable object was found in the
mass interval $0.67 M_0$ to $M_0$. Tests carried out with better
number statistics, involving a sub--sample size of 180 objects  and
requiring a minimum of nine detections in the lowest mass interval,
demonstrated that this procedure was robust.

Figure~\ref{fig:obsdistance} shows the distance out to which
simulated CHVCs of a given \hi mass can be detected with the HIPASS
survey. Both the \hi linewidth and spatial FWHM are dependent on the
dark--matter fraction, as illustrated in Fig.~\ref{fig:chvcmodel}, so
separate curves are shown for $\beta = -1.2$, $-1.6$, and
$-2.0$. A limiting case is provided by $\beta = -1.2$ which is
extremely dark--matter dominated for low \hi mass. In this case the
objects are so spatially extended (about 5~kpc FWHM) and have such a
high linewidth (about 60~\kms~FWHM) that they fall below the HIPASS
detection threshold for  log($M_{\rm HI}) < \sim6.4$. More plausible
linewidths and spatial FWHM follow for $\beta = -1.6$ and
$-2.0$. Such objects are sufficiently concentrated that they can still
be detected, even when highly resolved in the HIPASS data.

The clouds are regarded stable against ram--pressure stripping if the
gas pressure at the center of the cloud exceeds the ram pressure,
$P_{\rm ram} = n_{\rm halo} \cdot V^2$, for a cloud moving with
velocity~$V$ through a gaseous halo with density $n_{\rm halo}$.
Because both the gaseous halo density and the cloud velocity are the
highest if the distance of the cloud to the galaxy is the smallest, the
stability against ram--pressure stripping was evaluated at closest
approach.  We therefore kept track of the closest approach of each
test--particle to the Galaxy and to M~31, while simulating the velocity
field of the Local Group. We used a density profile for the Galactic
halo which is an adaptation of the model of Pietz et~al.
(\cite{pietz}), derived to explain the diffuse soft X--ray emission as
observed by ROSAT.  Whereas their model is flattened towards the
Galactic plane, we simply use a spherical density distribution, in
which the radial profile is equal to the Galactic plane density profile
of Pietz et al. The density at a distance $r$ from the Galactic center
is given by
\begin{equation}
    n(r) = n_0 \cdot \left(
        \frac{\cosh(r_\odot / h)} {\cosh(r / h)}
    \right)^2,
\end{equation}
where $n_0=0.0013\,{\rm cm}^{-3}$~is the central density, $h=12.5$\,kpc
is the scalelength of the distribution, and $r_\odot=8.5$\,kpc is the
radius of the solar orbit around the Galactic center.  According to
this model, the total mass in the Galactic halo is
$1.5\times10^9\rm\;M_\odot$.  To describe the halo around M~31, we use
the same expression and the same parameter values except for~$n_0$, for
which we use a value twice the Galactic one. Figure~\ref{fig:ramdistance}
shows an example of the calculated distance at which a cloud of a
particular \hi mass will be stripped. The clouds in this example are
assumed to have a relative velocity of~$200\rm\;km\;s^{-1}$ with
respect to the Galactic halo.

A cloud will be tidally disrupted if the gravitational tidal field of
either the Galaxy or M~31 exceeds the self--gravity of the cloud.
We consider a cloud stable if
\begin{equation} \label{eq:tidal}
    \frac{M_{\rm dark}}{\sigma^2} \ge
    \left| \frac{\rm d^2}{{\rm d} r^2} \Phi_{\rm iso} (r) \right|
    \cdot \sigma,
\end{equation}
where $\sigma$~is the spatial dispersion of the Gaussian describing the \hi
distribution in the cloud, $M_{\rm dark}(r\le\sigma)$ is the core mass of
the dark matter halo, and $|{\rm d}^2 \Phi_{\rm iso} (r) / {\rm d} r^2|$
is the tidal force of either the Galaxy or M~31.  Solving the equation for
$r$ shows that only the least massive clouds with the lowest dark--matter
fractions are likely to suffer from tidal disruption.  If the slope of the
\hi mass distribution is as steep as $-$2.0, then clouds with an \hi mass
less than~$10^5\rm\;M_\odot$ are tidally disrupted at distances of about
60~kpc, as shown in Fig.~\ref{fig:tidalfield}.  For $M_{\rm
HI} > 2\times10^5\rm\;M_\odot$, or $\beta~>~-2$, the clouds are stable.
Changing the radius at which Eq.~\ref{eq:tidal} is evaluated
from~$1 \sigma$ to~$2 \sigma$, does not dramatically change this
result.

\subsection{Results of the Local Group simulations}

Before searching for a global best fit, we determined the range of
parameter values over which at least a moderately good representation
of the observed data was possible with the simulated data.  To quantify
the degree of agreement between the simulated size--, column--density
and velocity--FWHM distributions with the observations, we used a
$\chi^2$--test taken from \S\,14.3 of {\it Numerical Recipes} (Press et
al. \cite{press}).  The size--, column--density, and velocity--FWHM
distributions of a simulation were considered reasonable if $\chi^2
({\rm size})<3$, $\chi^2 ({\rm N_{HI}})<5$, and $\chi^2 ({\rm
FWHM})<5$. The incorporation of the spatial and kinematic information
was done by comparing the modeled $(l,b)$, $(l,V_{\rm GSR})$, and
$(V_{\rm GSR},b)$ distributions with those observed.  We used the
two-dimensional K--S test described in \S\,14.7 of {\it Numerical
Recipes} to make this comparison. The fits were considered acceptable if
$\tilde\chi^2 (l,b)$, $\tilde\chi^2 (l,V_{\rm GSR})$, and $\tilde\chi^2
(V_{\rm GSR},b)$ where all less than 0.3.

Table~\ref{table:model} shows which part of the parameter space
produces moderately good fits. The best fits have a Gaussian dispersion
between $150$ and $250\rm\;kpc$, an upper \hi mass cut-off between
$10^{6.5}$ and $10^{8.0}\rm\;M_\odot$ and a slope of the \hi mass
distribution of~$-1.7$ to $-1.9$.

\begin{table*}
\def\t#1{$10^{#1}\rm\;M_\odot$}
\def\d{$\;\ldots\;$}
\def\x{\emptybox{999\d999}}

\caption{Tabulation of the range of parameters entering the Local Group
models described in \S\,\ref{sec:model}.  Three free parameters were
specified for each simulation, namely the maximum allowed \hi mass,
$M_1$ (in units of M$_\odot$), the slope of the \hi mass distribution,
$\beta$, and the Gaussian dispersion of the cloud population, $\sigma_d$ (in
units of kpc).  The simulated CHVC populations were subjected to the
same observational constraints as pertain to the LDS and HIPASS
surveys.  The simulated spatial and kinematic deployments, as well as
the simulated size--, column density-- and velocity--FWHM
distributions, were compared with those observed.  The table shows
which Gaussian density dispersions produce acceptable results for each
of the combinations of $M_1$ and $\beta$.  A field is blank if the
simulation returned no acceptable fit for the given parameter
combinations. Fits were deemed acceptable if $\chi^2{\rm size} < 3$,
$\chi^2{\rm N_{HI}} < 5$, $\chi^2{\rm FWHM} < 5$,
$\tilde\chi^2{(l,b)} < 0.3$, $\tilde\chi^2{(l,V_{\rm GSR})} < 0.3$,
and $\tilde\chi^2{(V_{\rm GSR},b)} < 0.3$.  }\label{table:model}

\begin{tabular}{r|ccccccc}
\hline
       & $M_1=$\t{6} & \t{6.5} & \t{7} & \t{7.5} & \t{8} & \t{8.5}  &
\t{9} \\
\hline 
$\beta=-1.2$ &    &         &            &            &            &
&            \\
$-1.4$ &          &            &            &            &            &
&            \\  
$-1.6$ &            &            &            &            &            &
&             \\
$-1.7$ &            &$\sigma_d$=150\d200 kpc&150\d250&150\d200& 150\d200&
&              \\
$-1.8$ &            &            &150\d250&150\d250&150\d250&
&            \\
$-1.9$ &            &            &            &200\d250&200\d250&
&            \\
$-2.0$ &            &            &            &            &            &
&            \\
\hline
\end{tabular}
\end{table*}

Each simulation contains a relatively small number of detectable
objects, namely the same number of objects as in our all--sky CHVC
catalog.  Therefore the $\chi^2$ values are prone to shot--noise.  By
performing a larger number of simulations for a specific combination of
free parameters, we are able to determine a more representative value
of $\chi^2$ for each model.  The most promising combinations of
parameter values, i.e. the entries in Table~\ref{table:model}, were
repeated 35~times to reduce the shot--noise, and the average results and
their dispersions are shown in Table~\ref{table:bestfit}.  The range of
resulting fit quality due purely to this shot--noise is illustrated in
Fig.~\ref{fig:qualmodel}, which shows model data with the highest and
lowest $\chi^2$ values from a sequence of 35 simulations.  

The best overall fits are fairly well--constrained to lie between
$\sigma_d=150$ and $200\rm\;kpc$, with an upper \hi mass cut--off of about
$10^{7}$ to $10^{7.5}\rm\;M_\odot$ and a slope of the \hi mass
distribution of~$-1.7$ to $-1.8$. Comparison with
Fig.~\ref{fig:chvcmodel} suggests that populations of these types have
sufficiently high peak \hi column densities that they can provide
self-shielding to the extragalactic ionizing radiation field for
M$_{\rm HI}~>~10^{5.5}\rm\;M_\odot$.  The results of simulations \#~9
and\#~3 from Table~\ref{table:bestfit} are shown in
Fig.~\ref{fig:model01} and Fig.~\ref{fig:model02}, respectively. A
single instance of each simulation has been used in the subsequent
figures that had $\chi^2$ values consistent with the ensemble average.

\begin{table*}
\caption{Average chi--square values and their dispersions for the 35
runs which were performed for each model of the Local Group deployment
of CHVCs.  Each model was specified by the indicated three parameters,
namely the Gaussian dispersion of the cloud population, $\sigma_d$, the
maximum allowed \hi mass, $M_1$, and the slope of the \hi mass
distribution, $\beta$.  Multiple runs yielded better estimates of each
$\chi^2$, reducing the sensitivity to the relatively small number of
objects in each individual simulation.  The output of each model was
sampled with the observational parameters of the LDS and HIPASS surveys
and compared with the observed data.
}\label{table:bestfit}

\begin{tabular}{c|ccccccccc}
\hline
\vspace{-.05cm} Model & $\sigma_d$  & $M_1$ & $\beta$ &
$\chi^2({\rm size})$  & $\chi^2({\rm N_{HI}})$  & $\chi^2({\rm FWHM})$  &
$\tilde\chi^2(l,b)$ & $\tilde\chi^2(l,V{\rm GSR})$  &  $\tilde\chi^2(V{\rm
GSR},b)$  \\ 
%\hline
  \# & (kpc) & $({\rm M}_\odot$)       &   
&      &      &      &
\\
\hline
 1 & 150 & $10^{6.5}$ & $-1.7$ & 2.0$\pm$0.6 & 1.9$\pm$0.3 &4.0$\pm$0.4 &
0.29$\pm$0.03 & 0.25$\pm$0.02 & 0.29$\pm$0.03 \\
 2 & 150 & $10^{7.0}$ & $-1.7$ & 2.2$\pm$0.5 & 2.3$\pm$0.4 &3.5$\pm$0.4 &
0.27$\pm$0.03 & 0.23$\pm$0.03 & 0.28$\pm$0.03 \\
 3 & 150 & $10^{7.5}$ & $-1.7$ & 2.6$\pm$0.6 & 2.7$\pm$0.4 &3.9$\pm$0.6 &
0.25$\pm$0.04 & 0.21$\pm$0.03 & 0.25$\pm$0.03 \\
 4 & 150 & $10^{8.0}$ & $-1.7$ & 3.0$\pm$0.6 & 3.0$\pm$0.5 &3.9$\pm$0.5 &
0.24$\pm$0.03 & 0.21$\pm$0.03 & 0.24$\pm$0.02 \\
 5 & 150 & $10^{7.0}$ & $-1.8$ & 2.0$\pm$0.5 & 2.4$\pm$0.5 &4.5$\pm$0.6 &
0.27$\pm$0.04 & 0.23$\pm$0.02 & 0.28$\pm$0.03 \\
 6 & 150 & $10^{7.5}$ & $-1.8$ & 2.4$\pm$0.5 & 2.6$\pm$0.5 &4.1$\pm$0.5 &
0.25$\pm$0.04 & 0.21$\pm$0.03 & 0.24$\pm$0.02 \\
 7 & 150 & $10^{8.0}$ & $-1.8$ & 2.7$\pm$0.6 & 2.7$\pm$0.4 &4.0$\pm$0.6 &
0.24$\pm$0.03 & 0.21$\pm$0.03 & 0.24$\pm$0.03 \\

 8 & 200 & $10^{6.5}$ & $-1.7$ & 2.4$\pm$0.5 & 2.9$\pm$0.3 &4.9$\pm$0.6 &
0.29$\pm$0.04 & 0.28$\pm$0.02 & 0.30$\pm$0.02 \\
 9 & 200 & $10^{7.0}$ & $-1.7$ & 2.3$\pm$0.5 & 2.9$\pm$0.3 &4.1$\pm$0.5 &
0.28$\pm$0.04 & 0.24$\pm$0.02 & 0.29$\pm$0.03 \\
10 & 200 & $10^{7.5}$ & $-1.7$ & 2.9$\pm$0.6 & 3.0$\pm$0.5 &4.8$\pm$0.7 &
0.26$\pm$0.04 & 0.22$\pm$0.03 & 0.26$\pm$0.03 \\
11 & 200 & $10^{8.0}$ & $-1.7$ & 3.0$\pm$0.5 & 3.1$\pm$0.5 &4.9$\pm$0.6 &
0.26$\pm$0.04 & 0.22$\pm$0.03 & 0.25$\pm$0.03 \\
12 & 200 & $10^{7.0}$ & $-1.8$ & 2.2$\pm$0.5 & 3.6$\pm$0.5 &4.2$\pm$0.5 &
0.28$\pm$0.04 & 0.24$\pm$0.03 & 0.29$\pm$0.03 \\
13 & 200 & $10^{7.5}$ & $-1.8$ & 2.5$\pm$0.6 & 3.7$\pm$0.4 &3.8$\pm$0.5 &
0.27$\pm$0.04 & 0.22$\pm$0.03 & 0.27$\pm$0.03 \\
14 & 200 & $10^{8.0}$ & $-1.8$ & 2.7$\pm$0.5 & 3.7$\pm$0.5 &3.6$\pm$0.5 &
0.26$\pm$0.04 & 0.22$\pm$0.02 & 0.26$\pm$0.04 \\
15 & 200 & $10^{7.5}$ & $-1.9$ & 2.3$\pm$0.5 & 4.0$\pm$0.5 &4.8$\pm$0.7 &
0.26$\pm$0.04 & 0.22$\pm$0.03 & 0.27$\pm$0.04 \\
16 & 200 & $10^{8.0}$ & $-1.9$ & 2.4$\pm$0.5 & 4.0$\pm$0.5 &4.3$\pm$0.8 &
0.26$\pm$0.05 & 0.22$\pm$0.02 & 0.26$\pm$0.04 \\

17 & 250 & $10^{7.0}$ & $-1.7$ & 2.3$\pm$0.6 & 4.3$\pm$0.6 &4.8$\pm$0.7 &
0.28$\pm$0.04 & 0.24$\pm$0.02 & 0.29$\pm$0.03 \\
18 & 250 & $10^{7.0}$ & $-1.8$ & 2.2$\pm$0.5 & 5.2$\pm$0.6 &4.1$\pm$0.5 &
0.26$\pm$0.04 & 0.23$\pm$0.02 & 0.27$\pm$0.03 \\
19 & 250 & $10^{7.5}$ & $-1.8$ & 2.5$\pm$0.6 & 5.1$\pm$0.6 &3.8$\pm$0.5 &
0.25$\pm$0.03 & 0.22$\pm$0.02 & 0.26$\pm$0.03 \\
20 & 250 & $10^{8.0}$ & $-1.8$ & 2.7$\pm$0.4 & 5.3$\pm$0.7 &3.6$\pm$0.6 &
0.25$\pm$0.04 & 0.22$\pm$0.02 & 0.25$\pm$0.03 \\
21 & 250 & $10^{7.5}$ & $-1.9$ & 2.3$\pm$0.5 & 5.6$\pm$0.6 &4.8$\pm$0.6 &
0.26$\pm$0.03 & 0.22$\pm$0.02 & 0.26$\pm$0.03 \\
22 & 250 & $10^{8.0}$ & $-1.9$ & 2.5$\pm$0.6 & 5.6$\pm$0.6 &4.3$\pm$0.6 &
0.25$\pm$0.03 & 0.22$\pm$0.02 & 0.25$\pm$0.03 \\
\hline
\end{tabular}
\end{table*}

Both of these Local Group simulations succeed reasonably well in
reproducing the observed kinematic and population characteristics of
the CHVC sample as summarized in Fig.~\ref{fig:dataoverview}. The CHVC
concentrations, named Groups~2, 3, and 4 above, while not reproduced in
detail, have counterparts in the simulations which arise from the
combination of Gaussian density distributions centered on the Galaxy
and M\,31, together with the Local Group velocity field, population
decimation by disruption effects, and the foreground \hi obscuration.
A notable success of these simulations is their good reproduction
of the smoothed velocity field, including both the numerical values and
the location of minima and maxima.

One aspect of the observed CHVC distributions which can not be
reproduced accurately by our simulations is the distribution of observed
linewidth. The model objects are assumed to contain only the warm
component of \hi with a minimum linewidth corresponding to a 8000~K
gas. The profiles are then further broadened by the contribution of
rotation as indicated in Eq.~\ref{eq:temperature}. The actual objects
are known to have cool core components (e.g. Braun and Burton
\cite{braun01}, Burton et al. \cite{burt01}) which can contribute a
significant fraction of the \hi mass and consequently lead to narrower
observed line profiles. This shortcoming of the model distributions is
illustrated in the central panel of Fig.~\ref{fig:qualmodel}. The
narrow--linewidth tail of the observed distributions can never be
reproduced by the models. The best--fitting models can only succeed in
reproducing the median value and high-velocity tail of the
distribution. 

In order to better isolate the effect of foreground obscuration from the
intrinsic distribution properties of the simulation itself we show the
unobscured version of model~\#9 in Fig.~\ref{fig:model01no}. Comparison
of Figs.~\ref{fig:model01} and~\ref{fig:model01no} illustrates how the
foreground obscuration from the \hi Zone of Avoidance  modifies the
distribution of object density. The location of apparent object
concentrations are shifted and density contrasts are enhanced.  The
comparison also reveals that the large gradient in the smoothed GSR
velocity field is an intrinsic property of the Local Group model and not
simply an artifact of the Galactic obscuration.  Substantial negative
velocities ($<-100$~\kms) are predicted in the direction of M\,31 (which
effectively defines the Local Group barycenter), while slightly positive
velocities are predicted in the anti-barycenter direction, just as
observed.

A better appreciation of the physical appearance of these Local Group
models is provided in Fig.~\ref{fig:mod3d}, where two perpendicular
projections of the model~\#\,9 population are displayed. The $(x,y)$
plane in the figure is the extended Galactic plane, with the Galaxy
centered at $(x,y)=(0,0)$ with the positive $z$ axis corresponding to
positive $b$. The intrinsic distribution of objects is an elongated cloud
encompassing both the Galaxy and M\,31, which is dominated in number by
the M\,31 concentration. The objects that have at some point in their
history approached so closely to either of these galaxies that their
\hi would not survive the ram--pressure or tidal stripping are indicated
by the filled black circles. Cloud disruption appears to have been
substantially more important in the M\,31 concentration than for the
Galaxy. The objects that are too faint to have been detected by the LDS
or HIPASS observations, depending on declination, are indicated by grey
circles. {\emph {The bulk of the M\,31 sub--concentration is not detected
in our CHVC sample for two reasons}}:~\,(1) these objects have a
larger average distance than the objects in the Galactic
sub--concentration, and (2)  the M\,31 sub--concentration is
located primarily in the northern celestial hemisphere, where the lower LDS
sensitivity compromises detection. We will return to this point below.

Those objects which are obscured by the \hi distribution of the Galaxy
are indicated in Fig.~\ref{fig:mod3d} by open red circles. Somewhat
counter--intuitively, the consequences of obscuration are not
concentrated toward the Galactic plane, but instead occur in the plane
perpendicular to the LGSR solar apex direction
$(l,b)=(93^\circ,-4^\circ)$.  This can be understood by referring back
to our discussion in \S\,\ref{sec:obscur} and the illustration in
Fig.~\ref{fig:calcskydistr}. Obscuration from the position-- and
velocity--dependent \hi Zone of Avoidance is most dramatic when the
kinematic properties of a population result in overlap with \vlsr$=0$
\kms, since this can occur over a large solid angle, while the Galactic
plane is relatively thin.

The various processes which influence the observed distributions are
further quantified in Table~\ref{table:modstat}. Matching the detected
sample size of 163 CHVCs above $b=-65^\circ$, required the simulation
of some 6300 objects in the case of models~\#\,9 and \#\,3. About three
quarters of the simulated populations were classified as disrupted due
to ram--pressure or tidal stripping; while 80\% of the remaining objects
were deemed too faint to detect with the LDS (in the north) or
HIPASS (in the south).  Obscuration by Galactic \hi eliminated about
one half of the otherwise detectable objects. The total \hi masses
involved in these two model populations were $4.3\times10^9$\,M$_\odot$
and $6.4\times10^9$\,M$_\odot$, respectively. In both cases, about 75\%
of this mass had already been consumed by M\,31 and the Galaxy via
cloud disruption, leaving only 25\% still in circulation, although
distributed over some 1200 low--mass objects.

\begin{table*}
\caption{Model statistics for the best--fit Local Group models.  The
total numbers of objects and their \hi masses are indicated for models
\#\,9 and \#\,3 (see Table~\ref{table:bestfit}) together with how these
are distributed within different categories. A detected sample of 163
objects was required in all cases after obscuration by the Galaxy and
excluding the anomalous south Galactic pole region at $b,-65\deg$. The
table also lists the number of objects classified as disrupted by
ram--pressure or tidal stripping as well as those too faint to be
detected by the LDS or HIPASS observations.  }\label{table:modstat}

\begin{tabular}{l|cccc}
\hline
Fate of input  CHVCs & \multicolumn{2}{c}{model \#9} &
\multicolumn{2}{c}{model \#3} \\
  & number & $M_{\rm HI}$  & number & $M_{\rm HI}$ \\
  & of CHVCs  & (10$^8$\,M$_\odot$) & of CHVCs & (10$^8$\,M$_\odot$) \\
\hline
Total number    & 6281 & 43  & 6310 & 64  \\
Disrupted by ram or tide &  4759 & 31 &  5178 & 50 \\
Too faint to be detected  &  1220 & 6.5 &  831 & 4.4  \\
          &      &     &      &     \\
Detectable if not obscured & 302 & 5.4 &  301 & 10 \\
          &      &     &      &     \\
Unobscured by \hi ZoA & 173 & 3.3 & 172 & 6.5 \\
Unobscured, not at SGP & 163 & 3.2 & 163 & 6.4 \\
\hline
\end{tabular}
\end{table*}

The crucial role of survey sensitivity in determining what is seen of
such Local Group cloud populations is also illustrated in
Figs.~\ref{fig:mod01glonv} and \ref{fig:mod01glatv}. The red symbols in
these figures indicate objects detectable with the relevant LDS or
HIPASS sensitivities, while the black symbols indicate those that
remain undetected due to either limited sensitivity or obscuration. If these
models describe the actual distribution of objects, then the prediction
is that future deeper surveys will detect large numbers of objects at
high negative LSR velocities in the general vicinity (about
$60\times60^\circ$) of M\,31. To make this prediction more specific, we
have imagined the sensitivity afforded by the current HIPASS survey in
the south extended to the entire northern hemisphere.
Fig.~\ref{fig:mod01hom} illustrates the prediction. A high
concentration of about 250 faint newly detected CHVCs is predicted in
the Local Group barycenter direction once HIPASS sensitivity is
available.

% ----------------------------------------------------------------------
% SIMPLE CHVC MODELS
% ----------------------------------------------------------------------

\section{A Galactic Halo population model for the CHVC
ensemble}
\label{sec:simplemodel}

In the previous section we have outlined a physical model for
self--gravitating, dark--matter dominated CHVCs evolving in the Local
Group potential. While that model was quite successful in describing
the global properties of the CHVC phenomenon, we noted that some
aspects of the observed kinematic and spatial deployment were strongly
influenced by the effects of obscuration by foreground Galactic \hi and
that, furthermore, the sensitivity limitations of the currently
available \hi survey material preclude tightly constraining the
characteristic distances.  In this section we consider to what extent a
straightforward model in which the CHVCs are distributed throughout an
extended halo centered on the Galaxy might also satisfy the
observational constraints.  We consider such a Galactic Halo model ad
hoc in the sense that it lacks the physical motivation that the
hierarchical structure paradigm affords the Local Group model.

We consider a spherically symmetric distribution of clouds,
centered on the Galaxy.  The radial density profile of the
population is described by a Gaussian function, with its peak located
at the Galactic center and its dispersion to be specified as a free
parameter of the simulations. The \hi mass distribution is given by a
power--law  the slope of which is a free parameter.  Different values are
allowed for the lowest \hi mass in the simulation.  The
\hi density distribution of an individual cloud is also described by a
Gaussian function.  The central volume density is the same for all clouds
in a particular simulation.  Given the \hi mass and central density of an
object, the spatial FWHM of the \hi distribution follows. For the
velocity FWHM we have simply adopted the thermal linewdth of an 8000~K
\hi gas of 21~\kms.

Each simulated cloud is ``observed'' with the parameters corresponding
to the LDS observations, if it is located in the northern celestial
hemisphere, but   with the HIPASS parameters if it is located in the
southern hemisphere.  Clouds are removed from the simulation if they are
too faint to be detected.  To include the effects of  obscuration by the
Milky Way, the velocity field of the clouds must be specified.  The
population is considered in the Galactic Standard of Rest system, where it
is distributed as a Gaussian with a mean velocity of
$-50\rm\;km\;s^{-1}$ and dispersion of $110\rm\;km\;s^{-1}$. These
values follow directly from the observed parameters summarized in
Table~\ref{table:velostat} after correction for obscuration as in
Fig.~\ref{fig:obscure}.  Clouds with a deviation velocity (as defined
in \S\,\ref{sec:obscur}) less than~$70\rm\;km\;s^{-1}$ are
removed. Additional clouds that pass the selection criteria are
simulated until their  number equals the number of  CHVCs
actually observed.

We  performed the simulations with the following values for the four
parameters that describe the distance, \hi mass, and spatial extent of
the population.
\begin{itemize}
\item The spatial dispersion of the cloud population. Values   range
from~$10\rm\;kpc$ to~$2\rm\;Mpc$; specifically we consider the values
of 10, 15, 20, \dots 50, 60, 70, \dots 100, 150, 200, \dots 500, 1000,
2000~kpc.
\item  The slope of the \hi mass distribution, $\beta$. Values for
$\beta$ were $-2.0, -1.8, \ldots, -0.8$.
\item   The lowest \hi mass, $M_0$, allowed in a simulation. Values for
$M_0$   were $10^2$, $10^3$, $10^4$, or~$10^5\rm\;M_\odot$.
\item  The central gas density in the clouds, $n_0$.  Values for $n_0$,
which remained constant for a single run, were $3\times10^{-3}$,
$1\times10^{-2}$, $3\times10^{-2}$, $0.1$, and~$0.3\rm\;cm^{-3}$.
\end{itemize}

The only measured quantities which can usefully be compared to the
models are the distributions of angular sizes and peak column densities.
This is because the average kinematics in these
simulations have already been defined to match the data by our choice
of the mean velocity and its dispersion. Given four free model
parameters for each simulation and only two distributions to determine
the degree of agreement between simulations and observations, it is
clear that the problem is under--determined.  We can only hope to
constrain the range of reasonable parameters in the four--dimensional
parameter space.

In order to assess the degree of agreement between the simulation outcomes
and the observations, we use a $\chi^2$--test from \S\,14.3 of {\it
Numerical Recipes}, (Press et al. \cite{press}).  The size and column
density distributions of the models and the data are compared.  A
simulation was considered acceptable if  $\chi^2 ({\rm size}) < 5$ and
$\chi^2 (N_{\rm HI}) < 5$.  Figure~\ref{fig:qualsimplemodel} shows
examples of the range of fit quality that was deemed acceptable for both
the column density and size distributions.

Table~\ref{table:simplemodel} lists the parameter combinations
that produce formally acceptable results, and shows that for each~$M_0$
value the acceptable solutions are concentrated around a line.  The
solutions range from nearby models, for which the central density is of the
order of~$0.1\rm\;cm^{-3}$, the mass slope is~$-2.0$, and the
characteristic distance is several tens of kpc, to more distant models,
having a central density of~$0.01\rm\;cm^{-3}$,   a mass slope of~$-1.4$,
and characteristic distances of several hundreds of kpc. Since column
density is simply the product of depth and density this coupling of
distance to central density is easily understood.

\begin{table*}
\def\ten#1{\cdot10^{#1}\rm\;cm^{-3}}
\def\d{$\;\ldots\;$}
\def\x{\emptybox{999\d999}}
\caption{Results of the models described in
\S\,\ref{sec:simplemodel}, in which the CHVCs are viewed as forming an
extended halo population, centered on the Galaxy.  Each simulation is
determined by four free parameters, namely the central \hi density of
the clouds, $n_0$, the slope of the \hi mass distribution, $\beta$, the
lowest \hi mass in each simulation, $M_0$, and the dispersion of the
spatial Gaussian that defines the distance,
$\sigma_d$, of the cloud population.  The simulations were sampled with the
observational parameters of the LDS and HIPASS surveys and compared
with the CHVC sample. The table shows which distance dispersions
produce acceptable results for each of the combinations of $n_0$,
$\beta$, and $M_0$.  A field is blank if there is no good fit for the
given parameter combinations.  A simulation is considered successful if
$\chi^2_{\rm size} < 5$ and $\chi^2_{\rm N_{HI}} < 5$.  The central
cloud density, $n_0$, is shown horizontally above each section of the
table, in units cm$^{-3}$; the slope of the \hi mass distribution is
listed vertically on the left of the table: $M_0$ ranges from
$10^2\rm\;M_\odot$ for the top table to $10^5\rm\;M_\odot$ for the
bottom table.  Distances are in kpc.  The model is not tightly
constrained, because of degeneracies in parameter combinations.
}\label{table:simplemodel}

Simulations with $M_0=10^2$:

\begin{tabular}{r|ccccc}
\hline
  & $n_0=3\ten{-3}$ & $1\ten{-2}$ & $3\ten{-2}$ & $1\ten{-1}$
&
$3\ten{-1}$
\\
\hline
%$\beta=-0.8$ & \x   & \x    & \x   & \x    &\x     \\  \hline
$\beta=-1.0$ & \x          & \x          & \x          & \x          &
\x         
\\  
$-1.2$ & $\sigma_d=$150\d300\,kpc    & 35\d45      & 10\d20  & \x  &\x
\\  
$-1.4$ & 200\d350    & 45\d100     & 15\d40      & 15          & \x
\\  
$-1.6$ & 200\d350    & \x          & \x          & \x          & \x
\\  
$-1.8$ & 500         & \x          & \x          & \x          & \x
\\  
$-2.0$ & \x          & \x          & \x          & \x          & \x
\\  
\hline
\end{tabular}

\bigskip
Simulations with $M_0=10^3$:

\begin{tabular}{r|ccccc}
\hline
  & $n_0=3\ten{-3}$ & $1\ten{-2}$ & $3\ten{-2}$ & $1\ten{-1}$ &
$3\ten{-1}$
\\
\hline
%$\beta=-0.8$ & \x   & \x    & \x   & \x    &\x     \\  \hline
$\beta=-1.0$ & \x          & \x          & \x          & \x          &
\x         
\\  
$-1.2$ & \x          & 70\d100     & \x          & \x          & \x
\\  
$-1.4$ & $\sigma_d=$100\d400\,kpc    & 45\d100     & 15\d40      & 15          &
\x         
\\  
$-1.6$ & 200         & 50, 90      & 15\d25      & 10\d15      & \x
\\  
$-1.8$ & \x          & \x          & 10\d15      & 10\d15      & \x
\\  
$-2.0$ & \x          & \x          & 10          & 10\d15      & \x
\\  
\hline
\end{tabular}

\bigskip
Simulations with $M_0=10^4$:

\begin{tabular}{r|ccccc}

\hline
   & $n_0=3\ten{-3}$ & $1\ten{-2}$ & $3\ten{-2}$ & $1\ten{-1}$ &
$3\ten{-1}$
\\
\hline
%$\beta=-0.8$ & \x   & \x    & \x   & \x    &\x     \\  \hline
$\beta=-1.0$ & \x          & \x          & \x          & \x          &
\x         
\\ 
$-1.2$ & $\sigma_d=$100\d250\,kpc   & \x          & \x          & \x          &
\x         
\\  
$-1.4$ & 150\d450    & 50\d100     & 50\d70      & \x          & \x
\\  
$-1.6$ & 400         & 45\d100     & 40\d50      & \x          & \x
\\  
$-1.8$ & \x          & 40\d60      & 30\d45      & \x          & \x
\\  
$-2.0$ & \x          & 40\d45      & 30\d45      & \x          & \x
\\  
\hline
\end{tabular}

\bigskip
Simulations with $M_0=10^5$:

\begin{tabular}{r|ccccc}
\hline
  & $n_0=3\ten{-3}$ & $1\ten{-2}$ & $3\ten{-2}$ & $1\ten{-1}$ &
$3\ten{-1}$
\\
\hline
%$\beta=-0.8$ & \x   & \x    & \x   & \x    &\x     \\  \hline
$\beta=-1.0$ & \x          & \x          & \x          & \x          &
\x         
\\ 
$-1.2$ & $\sigma_d=250$\,kpc         & \x          & \x          & \x          &
\x         
\\  
$-1.4$ & 250\d450    & 150         & \x          & \x          & \x
\\  
$-1.6$ & 150\d300    & 100\d150    & \x          & \x          & \x
\\  
$-1.8$ & 150\d200    & 100\d150    & \x          & \x          & \x
\\  
$-2.0$ & \x          & 90\d150     & \x          & \x          & \x
\\  \hline
\end{tabular}
\end{table*}

Overviews of two of the best--fitting models of the Galactic Halo
type are given in Figs.~\ref{fig:abestsimplemodel} and
\ref{fig:bbestsimplemodel}.  Figure~\ref{fig:abestsimplemodel} shows a
cloud population with 30~kpc dispersion, while the population in
Fig.~\ref{fig:bbestsimplemodel} has a dispersion of 200~kpc. These
figures can be compared with Fig.~\ref{fig:dataoverview}, showing the
situation actually observed.  Despite there being almost a factor of
ten difference in the average object distance for these two models,
they produce similar distributions of observables, which are to a large
extent determined by the effects of obscuration.  Relative to the
observed CHVC sample shown in Fig.~\ref{fig:dataoverview}, the density
distributions of these models are more uniformly distributed on the
sky. The average velocity fields are also more symmetric about
$ b=0\deg $, lacking the extreme negative excursion toward
$(l,b)=(125^\circ,-30^\circ)$ seen in the CHVC population, that
produces a large gradient in the $(V_{\rm GSR},b)$ plot.

% ----------------------------------------------------------------------
% DISCUSSION
% ----------------------------------------------------------------------
\section{Discussion and conclusions}\label{sec:conclusion}

The effects of both obscuration by the gaseous disk of the Galaxy and the
limited sensitivity of currently available \hi  surveys have
important consequences for the observed properties of the HVC phenomenon.
We have identified those consequences in this paper. Obscuration leads to
apparent localized enhancements of object density, as well as to
systematic kinematic trends that need not be inherent to the population
of CHVCs. A varying resolution and sensitivity over the sky substantially
complicates the interpretation of the observed distributions. Taking
account of both these effects in a realistic manner is crucial to
assessing the viability of models for the origin and deployment of the
anomalous--velocity H\,{\sc i}.  Our discussion leads to specific
predictions for the numbers and kinematics of faint CHVCs which can be
tested in future \hi surveys.

\subsection{Galactic Halo models}
As shown in \S\,\ref{sec:simplemodel}, a straightforward empirical
model in which CHVCs are dispersed throughout an extended halo centered on
the Galaxy does not provide the means to discriminate between distances
typical of the Galactic Halo and those of the Local Group. Comparable fit
quality is realized for distance dispersions ranging from about 30 to
300~kpc. In addition to requiring a relatively large number of free
parameters, such empirical models beg a number of serious physical
questions. In the first instance: how is it that \hi clouds can survive at
all in a low--pressure, high--radiation--density environment without the
pressure support given by a dark halo? Presumably such ``naked'' Galactic
Halo \hi clouds would either be very short--lived or require continuous
replenishment, since the timescales for reaching thermal and pressure
equilibrium are only about 10$^7$~years (Wolfire et al.  \cite{wolf95}).
Realistic assessment of such a scenario must await more detailed
simulations that track the long--term fate of gas, for example after tidal
stripping from the LMC/SMC, within the Galactic Halo.  Only by including
more physics will it be possible to reduce the number of free parameters
and determine meaningful constraints on this type of scenario. This class
of model also suffers from a number of shortcomings in describing the
observed distributions, namely that the object density enhancement coupled
with high negative velocities seen in the Local Group barycenter direction
are not reproduced.

The Galactic Halo simulations returned formally acceptable values of
characteristic distance as low as some 30 kpc.  There is, however, a
growing body of independent evidence based on high--resolution imaging
of a limited number of individual CHVCs that such nearby distances do
not apply. Braun \& Burton (\cite{braun00}) discussed evidence from
Westerbork synthesis observations of rotating cores in
?HVC\,204.2\,$+$\,29.8\,$+$\,075 (using the de\,Heij et al.
\cite{deheij02} notation for a semi-isolated source) whose internal
kinematics could be well modeled by rotation curves in flattened disk
systems within cold dark matter halos as parameterized by Navarro et
al. (\cite{navarro97}), if at a distance of at least several hundred
kpc. Similar distances were indicated for
?HVC\,115.4\,$+$\,13.4\,$-$\,260 on the basis of dynamical stability
and crossing--time arguments regarding the several cores observed with
different systemic velocities, but embedded in a common diffuse
envelope.  The WSRT data for CHVC\,125.3\,$+$\,41.3\,$-$\,205 likewise
supported distances of several hundred kpc, based on a volume--density
constraint stemming from the observed upper limit to the kinetic
temperature of 85\,K.  Burton et al. (\cite{burt01}) found evidence in
Arecibo imaging of ten CHVCs for exponential edge profiles of the
individual objects: the outer envelopes of the CHVCs are not tidally
truncated and thus are likely to lie at substantial distances from the
Milky Way.  For plausible values of the thermal pressure at the
core/halo interface, these edge profiles support distance estimates
which range between 150 and 850 kpc.

\subsection{Local Group models}
The Local Group deployment models of \S\,\ref{sec:model} offer a more
self--consistent and physically motivated scenario for the CHVC
population.  Dark--matter halos provide the gravitational confinement
needed to produce a two--phase atomic medium with cool \hi
condensations within warm \hi envelopes, and provide in addition the
necessary protection against ram--pressure and tidal stripping to allow
long--term survival. The kinematics of the population follow directly
from an assumed passive evolution within the Local Group potential.
While three free parameters (the distance scalelength, the mass
function slope, and the upper mass cut--off) were then tuned to explore
consistency with the observations, only the distance was effectively a
``free'' parameter. The mass function slopes of the best fits have
values of~$-1.7$ to $-1.8$, in rough agreement with the value of
$-1.6$ favored by Chiu et al. (\cite{chiu}) for the distribution of the
baryonic masses in their cosmological simulations. The somewhat steeper
slopes and therefore larger baryonic fractions favored by our
model fits might be accomodated by recondensation onto the dark--matter
halos at later times. 

The \hi upper mass cut--off introduced in the Local Group models can
also be externally constrained.  In addition to satisfying the
observational demand that no \hi column densities exceeding a few times
10$^{20}$\,cm$^{-2}$ are seen in the CHVC population (consistent with
the absence of current internal star formation), there is the observed
lower limit of about 3$\times10^7$\,M$_\odot$ for the \hi mass seen in
a large sample of late--type dwarf galaxies (Swaters
\cite{swat99}). The upper mass cut--off favored by the simulations, of
about $10^7$\,M$_\odot$, is essentially unavoidable given these two
constraints. 

The spatial Gaussian dispersion which is favored by these simulations
is quite tightly constrained to lie between about 150 and 200 kpc. The
implication for the distribution of object distances is illustrated in
Fig.~\ref{fig:mod01dist} in the form of a histogram of the detected
objects from model \#9. The distribution has a broad peak extending
from about 200 to 450~kpc with a few outliers extending out to 1~Mpc
due primarily to the M31 sub--population.  The filled circles in the
figure are the distance estimates for individual CHVCs found by Braun \&
Burton (\cite{braun00}) and Burton et al. (\cite{burt01}). Although very
few in number, these estimates appear consistent with the model
distribution, also peaking in number near 250~kpc.

We have made the simplifying assumption that the baryonic matter in our
model clouds is exclusively in the form of \hi, rather than being
partially ionized. It is reassuring that the best-fitting models have
peak column densities which are sufficiently high that the objects
should be self-shielding to the extragalactic ionizing radiation field
for M$_{\rm HI}~>~10^{5.5}$\,M$_\odot$ as noted above. Since the
neutral component requires a power--law slope of about $-1.7$ to fit
the data, it seems likely that the total baryonic mass distribution
might follow an even steeper distribution, since the mass fraction of
ionized gas will increase toward lower masses.

\subsection{The Local Group mass function}
An interesting question to consider is whether the extrapolated mass
distributions of our Local Group CHVC models can also account for the
number of galaxies currently seen. In Fig.~\ref{fig:lgall} we plot the
mass distribution of objects in one of the best--fitting Local Group
models, model \#9 of Table~\ref{table:bestfit}. The thin--line
histogram gives the mass distribution of the model population after
accounting for the effects of ram--pressure and tidal stripping. The
thick--line histogram gives the observed CHVC distribution that results
from applying the effects of Galactic obscuration and sensitivity
limitations appropriate to the LDS and HIPASS properties in the
northern and southern hemispheres, respectively. The hatched histogram
gives the inferred total baryonic (\hi plus stellar) mass distribution
of the Local Group galaxies tabulated by Mateo (\cite{mateo}), assuming
a stellar mass--to--light ratio of $M/L_B =
3$\,M$_\odot$/L$_\odot$. M31 and the Galaxy, with baryonic masses of
some 10$^{11}$\,M$_\odot$, are not included in the plot.  The diagonal
line in the figure has the slope of the model \hi mass function of
$\beta=-1.7$. The figure demonstrates that the low--mass populations of
these models are roughly in keeping with what is expected from the
number of massive galaxies together with a constant mass function slope
of about $\beta=-1.7$. At intermediate masses,
10$^7$--10$^{8.5}$\,M$_\odot$, there is a small deficit of cataloged
Local Group objects relative to this extrapolated distribution, while
at higher masses there is a small excess. Conceivably this may be the
result of galaxy evolution by mergers. 

It is important to note that the distribution of objects shown in
Fig.~\ref{fig:lgall} is only the current relic of a much more extensive
parent population. As shown in Table~\ref{table:modstat}, about 75\% of
the CHVC population in these models is predicted to have been disrupted
by ram pressure or tidal stripping over a Hubble time, contributing about
$3\times10^9$~M$\odot$ of baryons to the Local Group environment and
the major galaxies.

\subsection{The M\,31 population of CHVCs}
One of the most suggestive attributes of the CHVC population in favor
of a Local Group deployment is the modest concentration of objects
which are currently detected in the general direction of M\,31, i.e.
in the direction of the Local Group barycenter.  These objects have
extreme negative velocities in the GSR reference frame. While this is a
natural consequence of the Local Group models it does not follow from
the empirical Galactic halo models, nor is it a consequence of
obscuration by Galactic H\,{\sc i}. Putman \& Moore (\cite{putman02c})
have made some comparisons between numerical simulations of dark matter
mini--halos in the Local Group with the $(l,V_{\rm LGSR})$
distributions of HVCs and CHVCs, and were led to reject the possibility
of CHVC deployment throughout the Local Group. Our discussion here has
shown that such comparisons require taking explicit account of
detection thresholds in the available survey observations, as well as
of the vagaries of obscuration caused by the \hi Zone of Avoidance. The
Putman \& Moore investigation did not take these matters into account.
The modest apparent amplitude of the M\,31 concentration relative to
the Galactic population as seen with present survey sensitivities
provides the best current constraints on the global distance scale of
the CHVC ensemble. There follows a testable prediction, namely that
with increased sensitivity a larger fraction of the M\,31 population of
CHVCs should be detected. This prediction was made explicit in
Fig.~\ref{fig:mod01hom}, where one of our model distributions was
shown as it would have been detected if HIPASS sensitivity were
available in the northern sky. For that particular model, some 250
additional detected objects are predicted, of which the majority are
concentrated in the $60\times60^\circ$ region centered on M\,31. The
ongoing HIJASS survey of the sky north of $\delta=25^\circ$ (Kilborn
\cite{kilborn}), which is being carried out using the 76--m Lovell
Telescope at Jodrell Bank to about the same velocity coverage, angular
resolution, and sensitivity as the HIPASS effort, should allow this
prediction to be tested.

\subsection{The Sculptor Group lines of sight}
We have omitted the part of the sky around the south Galactic pole in
our fitting of Local Group models to the observations, because of the
extreme velocity dispersions measured in this direction. The
nearest external group of galaxies, the Sculptor Group, is located in
the direction of the south Galactic pole.  If the CHVCs are distributed
around the major Local Group galaxies, then plausibly the same sort of
objects could be present in the Sculptor Group.  Putman et al.
(\cite{putman02a}) mention detection of clouds in the direction of the
southern part of the Sculptor Group.  Because no similar clouds were
detected in the northern part of this Group, they consider it unlikely
that this concentration of CHVCs is associated with the Sculptor Group.
We note, however, that rather than being a spherical concentration of
galaxies, the Sculptor Group has an extended filamentary morphology,
which ranges in distance from~$1.7\rm\;Mpc$ in the south
to~$4.4\rm\;Mpc$ in the north.  Putman et al. assumed that the HIPASS
sensitivity would allow detection of \hi masses
of~$7\times10^6\rm\;M_\odot$ throughout the Sculptor Group.  But in
Fig.~\ref{fig:obsdistance} we show the actual distance out to which
HIPASS can detect \hi masses given a realistic cloud model and
detection threshold: even the most massive and rare objects in our
simulated distributions, with $M_{\rm HI}$~=~$10^7\rm\;M_\odot$, can
only be detected out to $2.5\rm\;Mpc$. It is therefore only the near
portion of the Sculptor filament that might be expected to show any
enhancement in CHVC density with the currently available sensitivities.

\subsection{Predicted CHVC populations in other galaxy groups}
It is also interesting to consider whether the simulated Local Group
model populations would be observable in external galaxy groups at even
larger distances. In Fig.~\ref{fig:mod01mass} we show one of our
best--fitting Local Group models, model \#9 of
Table~\ref{table:bestfit}, projected onto a plane as in
Fig.~\ref{fig:mod3d}. In Fig.~\ref{fig:mod3d}, the surviving
clouds were distinguished by \hi flux; in Fig.~\ref{fig:mod01mass},
the distinction is by \hi mass.  We indicate with grey dots those
objects that were deemed to have been disrupted by ram--pressure or
tidal stripping. The red and black dots indicate the remaining objects
in the population, with the red dots representing objects that exceed
$M_{\rm HI}$~=~3$\times10^6$\,M$_\odot$ and the black dots those that
fall below this mass limit. The choice of a limiting mass of $M_{\rm
HI}$=3$\times10^6$\,M$_\odot$ over a linewidth of 35~\kms~was made to
represent what might be possible for a deep \hi survey of an external
galaxy group. In this example, some 95 objects occur which exceed this
mass limit distributed over a region of some 1.5$\times$1.0 Mpc
extent. For a limiting mass of $M_{\rm HI}$=5$\times10^6$\,M$_\odot$
over 35~\kms, the number drops to 45. It is clear that a very good mass
sensitivity will be essential to detecting such potential CHVC
populations in external galaxy groups. Current searches for such
populations, reviewed by Braun \& Burton (\cite{braun01}), have
generally not reached a sensitivity as good as even $M_{\rm
HI}$=$10^7\rm\;M_\odot$ over 35~\kms, so it is no surprise that such
distant CHVCs have not yet been detected.

\begin{acknowledgements}
  The Westerbork Synthesis Radio
  Telescope is operated by the Netherlands Foundation for Research in
  Astronomy, under contract with the Netherlands Organization for
  Scientific Research.
\end{acknowledgements}


\begin{thebibliography}{}
\bibitem[1987]{baja87}
Bajaja, E., Morras, R., \& P\"oppel, W.\,G.\,L. 1987,
Pub. Astr. Inst. Czech. Ac. Sci., 69, 237
\bibitem[2001]{barnes}
Barnes, D.\,G., Staveley--Smith, L., de Blok, W.\,J.\,G., et al. 2001,
MNRAS, 322, 486
%\bibitem[1998]{bland98}
%    Bland--Hawthorn, J., Veilleux, S., Cecil, G.\,N., Putman, M.\,E.,
%Gibson, B.\,K., Maloney, P.\,R. 1998, MNRAS, 299, 611
\bibitem[1999]{blitz99}
    Blitz, L., Spergel, D.\,N., Teuben, P.\,J., Hartmann, D., \& Burton,
W.\,B.   1999, ApJ, 514, 818
\bibitem[1999]{braun99}
    Braun, R., \& Burton, W.\,B. 1999, A\&A, 341, 437
\bibitem[2000]{braun00}
    Braun, R., \& Burton, W.\,B. 2000, A\&A, 354, 853
\bibitem[2001]{braun01}
Braun, R., \& Burton, W.\,B. 2001, A\&A, 375, 219
\bibitem[1995]{burkert}
    Burkert, A. 1995, ApJ, 447, L25-L28
\bibitem[1980]{breg80}
Bregman, J.\,N. 1980, ApJ, 236, 577
\bibitem[1999]{burt99}
Burton, W.\,B., Braun, R., Walterbos, R.\,A.\,M. \& Hoopes,
C.\,G. 1999, AJ, 117, 194
\bibitem[2001]{burt01}
Burton, W.\,B., Braun, R., \& Chengalur, J.\,N. 2001, A\&A, 369, 616,
(erratum in A\&A, 375, 227)
\bibitem[2001]{chiu}
    Chiu, W.\,A., Gnedin, N.\,Y., \& Ostriker, J.\,P. 2001, ApJ, 563, 21
\bibitem[2002]{choi02}
Choi, P.\,I., Guhathakurta, P., \& Johnston, K.\,V. 2002, AJ, submitted;
see also astro--ph/0111465
\bibitem[1993]{corbelli93}
Corbelli E., Salpeter E.\,E., 1993, ApJ, 419, 104
\bibitem[2002]{deheij02}
de\,Heij, V., Braun, R., \& Burton, W.\,B. 2002, A\&A, in press, 
astro--ph/0201249; Paper~I
\bibitem[1976]{eich76}
Eichler, D. 1976, ApJ, 208, 694
\bibitem[1976]{eina76}
Einasto, J., Haud, U., J\^oeveer, M., \& Kaasik A. 1976, MNRAS, 177, 357
%\bibitem[1994]{ferr94}
%Ferrara, A. \& Field, G.\,B. 1994, ApJ, 423, 665
\bibitem[1981]{giov81}
Giovanelli, R. 1981, AJ, 86, 1468
\bibitem[1996]{hartmann96}
Hartmann, D., Kalberla, P.\,M.\,W., Burton, W.\,B., \& Mebold, U. 1996,
A\&AS, 119, 115
\bibitem[1997]{hartmann97}
    Hartmann, D., \& Burton, W.\,B. 1997,
    Atlas of Galactic Neutral Hydrogen, Cambridge University Press
\bibitem[1999]{helmi99}
Helmi, A., White, S.\,D.\,M., de Zeeuw, P.\,T., \& Zhao, H.\,S. 1999,
Nature, 402, 53 
\bibitem[1998]{henn98}
Henning, P.\,A., Kraan-Korteweg, R.\,C., Rivers, A.\,J., Loan, A.\,J.,
Lahav, O., Burton, W.\,B. 1998, AJ, 115, 584
\bibitem[2000]{henn00}
Henning, P.\,A., Staveley-Smith, L., Ekers, R.\,D., Green, A.\,J., Haynes,
R.\,F., et al. 2000, AJ, 119, 2686
\bibitem[1978]{hulsbosch78}
Hulsbosch, A.\,N.\,M. 1978, A\&A, 66, L5
\bibitem[1994]{ibat94}
Ibata, R., Gilmore, G., \& Irwin, M. 1994, Nature, 370, 194
\bibitem[2001]{ibat01}
Ibata, R., Irwin, M., Lewis, G., et al. 2001, Nature, 412, 49
\bibitem[1996]{karachentsev96}
    Karachentsev, I.\,D., \& Makarov, D.\,A.  1996, AJ, 111, 794
\bibitem[2000]{jura00}
Juraszek, S.\,J., Staveley-Smith, L., Kraan-Korteweg, R.\,C., Green, A.\,J.,
Ekers, R.\,D., et al. 2000, AJ, 119, 1627
\bibitem[2002]{kilborn}
Kilborn, V.\,A. 2002, in Seeing through the Dust, ASP Conf. Series, eds.
R. Taylor, T. Landecker, \& T. Willis, in press
\bibitem[1999]{klyp99}
Klypin, A., Kravtsov, A.\,V., Valenzuela, O., \& Prada, F. 1999, ApJ, 522,
82
\bibitem[1986]{kraan}
    Kraan--Korteweg, R.\,C. 1986, A\&AS, 66, 255
\bibitem[2000]{maje00}
Majewski, S., Ostheimer, J.\,C., Patterson, R.\,J., et al. 2000, AJ, 119,
760
\bibitem[1993]{maloney93}
Maloney P., 1993, ApJ, 414, 41
\bibitem[1998]{mateo}
    Mateo, M. L. 1998, ARA\&A, 36, 435
\bibitem[2001]{moor01}
Moore, B., Calcaneo--Roldan, C., Stadel, J., et al. 2001,
Phys. Rev. D., 64, 3508; astro--ph/0106271
\bibitem[1999]{moor99}
Moore, B., Ghigna, S., Governato, G., Lake, G., Quinn, T., Stadel, J.,
\& Tozzi, P. 1999, ApJ, 524, L19
\bibitem[1963]{muller63}
    Muller, C.\,A., Oort, J.\,H., \& Raimond, E. 1963,
    C. R. Acad. Sci. Paris 257, 1661
\bibitem[1997]{navarro97}
Navarro, J.\,F., Frenk, C.\,S., \& White, S.\,D.\,M. 1997, ApJ, 490, 493
\bibitem[1926]{oort26}
Oort, J.\,H. 1926, Ph.\,D. Theses, University of Groningen
\bibitem[1966]{oort66}
    Oort, J.\,H. 1966, Bull. Astr. Inst. Netherlands, 18, 421
\bibitem[1970]{oort70}
    Oort, J.\,H. 1970, A\&A, 7, 381
\bibitem[1981]{oort81}
    Oort, J.\,H. 1981, A\&A, 94, 359
\bibitem[1998]{pietz}
Pietz, J., Kerp, J., Kalberla, P.\,M.\,W., Burton, W.\,B., Hartmann, D.,
\& Mebold, U. 1998, A\&A, 332, 55
\bibitem[1993]{press}
Press, W.\,H., Teukolsky, S.\,A., Vetterling, W.\,T., \& Flannery, B.\,P.
1993, Numerical Recipes in C, Cambridge University Press
\bibitem[1999]{putm99}
Putman, M.\,E. \& Gibson, B.\,K. 1999, PASA, 16, 70
\bibitem[2000]{putman}
Putman, M. E. 2000, Ph.\,D. Thesis, Australian National University
\bibitem[2002]{putman02a}
    Putman, M.\,E., de\,Heij, V., Stavely--Smith, L., et al. 2002, AJ,
123, 873
\bibitem[2002]{putman02c}
Putman, M.\,E., \& Moore, B. 2002, in Extragalactic Gas at Low Redshift,
ASP Conf. Series, eds. J. Mulchaey \& J. Stocke, in press; astro-ph/0110417
%\bibitem[2002b]{putman02b}
%    Putman, M. E., Staveley--Smith, L., Freeman, K. C., et al. 2002b, in
%preparation
\bibitem[1989]{raychaudhury}
    Raychaudhury, S., \&  Lynden--Bell, D. 1989, MNRAS, 240, 115
\bibitem[1986]{sandage86}
    Sandage, A. 1986, ApJ, 307, 1
\bibitem[1976]{shap76}
Shapiro, P.\,R., \& Field, G.\,B. 1976, ApJ, 205,762
\bibitem[1999]{swat99}
Swaters, R.\,A.  1999, Ph.\,D. Theses, University of Groningen
\bibitem[1994]{berg94}van den Bergh, S.  1994, AJ 107, 1328
\bibitem[1999]{woer99}
van Woerden, H., Schwarz, U.\,J., Peletier, R.\,F., Wakker, B.\,P.,
Kalberla, P.\,M.\,W. 1999, Nature, 400, 138
\bibitem[1975]{vers75}
Verschuur, G.\,L. 1975, ARA\&A, 13, 257
\bibitem[1999]{voskes}
Voskes, T., \& Burton, W.\,B. 1999, in ASP Conf. Ser 168, New Perspectives
on the Interstellar Medium, ed. A.\,R. Taylor, T.\,L. Landecker, \& G.
Joncas, 375
\bibitem[1990]{wakker90}
    Wakker, B.\,P. 1990, Ph.\,D. Thesis, University of Groningen
\bibitem[2001]{wakker01}
Wakker, B.\,P. 2001, ApJS, 136, 463
%\bibitem[1991]{wakk91b}
%Wakker, B.\,P., \& Schwarz, U. 1991, A\&A, 250, 484
\bibitem[1991]{wakk91c}
Wakker, B.\,P., \& van Woerden, H. 1991, A\&A, 250, 509
\bibitem[1997]{wakker97}
    Wakker, B.\,P., \& van Woerden, H. 1997, ARA\&A, 35, 217
%\bibitem[1999]{wakker99}
%    Wakker, B.\,P., van Woerden, H., \& Gibson, B.\,K. 1999,
%    in ASP Conf. Ser. 166, Stromlo Workshop on High-Velocity Clouds,
%    eds. B. K. Gibson \& M.E. Putman, 311
\bibitem[1995]{wolf95}
Wolfire M.\,G., McKee C.\,F., Hollenbach D., Tielens A.\,G.\,G.\,M.
1995, ApJ, 453, 673
\end{thebibliography}
\end{document}